\documentclass[a4paper,11pt]{article}
\usepackage[dvipdfmx]{graphicx}
\usepackage{color}
\usepackage{mathrsfs}
\usepackage{bm}
\usepackage{multirow}
\usepackage{booktabs}
\usepackage{amsmath,amssymb}
\usepackage[left]{lineno}
\newcommand\ASTART{\bigskip\noindent\begin{minipage}[b]{0.5\linewidth}}

\newcommand\AENDSKIP{\end{minipage}\bigskip}
\newcommand\AEND{\end{minipage}}

\newcommand{\Part}[3]{ \frac{ \partial^{#3} #1 }{ \partial #2^{#3} } }
\newcommand{\V}[1]{\bm{#1} } 
\newcommand{\Tr}[1]{ \mathop{\rm Tr}_{ #1 } }

\newcommand{\Ave}[1]{\left\langle {#1} \right\rangle} 

\newcommand{\Extr}[1]{ \mathop{\rm Extr}_{ #1 } }

\newcommand{\mR}{\mathbb{R}}

\newcommand{\lb}{\left(}
\newcommand{\rb}{\right)}
\newcommand{\lbb}{\left\{}
\newcommand{\rbb}{\right\}}
\newcommand{\lsb}{ \left[ }
\newcommand{\rsb}{ \right] }

\newcommand{\T}[1]{\tilde{#1}}
\newcommand{\Req}[1]{eq.\ (\ref{eq:#1})}
\newcommand{\BReq}[1]{Eq.\ (\ref{eq:#1})}
\newcommand{\NReq}[1]{(\ref{eq:#1})}
\newcommand{\Reqs}[2]{eqs.\ (\ref{eq:#1},\ref{eq:#2})}

\newcommand{\Reqss}[2]{eqs.\ (\ref{eq:#1}-\ref{eq:#2})}

\newcommand{\Rfig}[1]{Fig.\ \ref{fig:#1}}
\newcommand{\Rfigs}[2]{Figs.\ \ref{fig:#1} and \ref{fig:#2}}

\newcommand{\Lfig}[1]{\label{fig:#1}}
\newcommand{\Leq}[1]{\label{eq:#1}}
\newcommand{\Rsec}[1]{sec.\ \ref{sec:#1}}
\newcommand{\Lsec}[1]{\label{sec:#1}}
\newcommand{\be}{\begin{eqnarray}}
\newcommand{\ee}{\end{eqnarray}}
\newcommand{\ba}{\begin{array}}
\newcommand{\ea}{\end{array}}
\newcommand{\no}{\nonumber}

\newcommand{\subbe}{\begin{subequations}}
\newcommand{\subee}{\end{subequations}}
\newcommand{\bs}{\backslash}
\newcommand{\mc}[1]{\mathcal{#1}}
\DeclareMathOperator*{\argmin}{arg\,min}
\DeclareMathOperator*{\argmax}{arg\,max}

\newcommand{\RSS}{\mathcal{E}}

\newcommand{\IT}{\beta}
\newcommand{\str}{K}
\newcommand{\cbar}{d}
\newcommand{\Bias}{\hat{b}}

\title{Learning performance in inverse Ising problems \\ with sparse teacher couplings}
\author{
Alia Abbara\thanks{Laboratoire de Physique de l'Ecole Normale Sup\'erieure, 24 rue Lhomond, 75005 Paris, France}, 
Yoshiyuki Kabashima\thanks{
Department of Mathematical and Computing Science, Tokyo Institute of Technology, 2-12-1, Ookayama, Meguro-ku, Tokyo, Japan},
Tomoyuki Obuchi\thanks{Department of Systems Science, Graduate School of Informatics, Kyoto University, Yoshida Hon-machi, Sakyo-ku, Kyoto-shi, Kyoto 606-8501, Japan}
\thanks{Corresponding author. E-mail: obuchi$@$i.kyoto-u.ac.jp}
, and Yingying Xu$^{\dag}$}

\begin{document}
\maketitle

\begin{abstract}
We investigate the learning performance of the pseudolikelihood maximization method for inverse Ising problems. In the teacher-student scenario under the assumption that the teacher's couplings are sparse and the student does not know the graphical structure, the learning curve and order parameters are assessed in the typical case using the replica and cavity methods from statistical mechanics. Our formulation is also applicable to a certain class of cost functions having locality; the standard likelihood does not belong to that class. The derived analytical formulas indicate that the perfect inference of the presence/absence of the teacher's couplings is possible in the thermodynamic limit taking the number of spins $N$ as infinity while keeping the dataset size $M$ proportional to $N$, as long as $\alpha=M/N > 2$. Meanwhile, the formulas also show that the estimated coupling values corresponding to the truly existing ones in the teacher tend to be overestimated in the absolute value, manifesting the presence of estimation bias. These results are considered to be exact in the thermodynamic limit on locally tree-like networks, such as the regular random or Erd\H{o}s--R\'enyi graphs. Numerical simulation results fully support the theoretical predictions. Additional biases in the estimators on loopy graphs are also discussed. 
\end{abstract}

\section{Introduction} \Lsec{Introduction}
Inference based on the classical Ising model is called the {\it inverse Ising problem} or {\it Boltzmann machine learning}, which is attracting more and more attention with the increasing interest in machine learning technologies. One recent application spurring this trend is for retinal neurons~\cite{Schneidman2006,Shlens2006}, and subsequent applications to a wide range of systems have been conducted~\cite{Tang2008,Hamilton13,Watanabe2013a,Watanabe2014,Tkacik2014a,Tavoni2015,terada2018objective,terada2018objective2}, showing the potential usefulness of the inverse Ising framework. 

However, it is difficult to compute standard estimators such as the maximum likelihood estimator in this framework when the system size is large. Thus, certain approximations and/or algorithms must be tailored to ease this difficulty and meet the demands of advanced applications, which have been attempted in previous studies~\cite{besag1974spatial,kappen1998efficient,tanaka1998mean,hinton2002training,broderick2007faster,sessak2009small,roudi2011mean,mezard2011exact,cocco2011adaptive,aurell2012inverse,
vuffray2016interaction,lokhov2018optimal,vuffray2019efficient}. One of the most effective examples is the pseudolikelihood method~\cite{besag1974spatial,aurell2012inverse}. This method approximates the likelihood function by the product of conditional likelihood functions, each of which is for a single random variable conditioned by its neighboring variables. This is useful for a wide range of problems defined on graphical models, and enables us to treat large systems because local couplings directly connected to a focused single random variable are isolated from the other couplings, and thus can be estimated independently. This local nature is sometimes referred to as {\it local learning}~\cite{bachschmid2017statistical}. 

Another benefit of local learning is its theoretical tractability in high-dimensional settings. Recent theoretical analyses based on the replica method revealed the tight limit of inference accuracy in the thermodynamic limit where the dataset size is comparable to model dimensionality~\cite{bachschmid2017statistical,bachschmid2015learning,berg2017statistical}. This provides a firm theoretical basis for the inverse Ising framework.  

Previous studies of~\cite{bachschmid2017statistical,bachschmid2015learning,berg2017statistical} focused on fully-connected Ising models. In high-dimensional settings, however, sparsely-connected models are more interesting because the inference accuracy is expected to be much better than the dense case, and the inferred estimator is expected to have clearer interpretations owing to the sparsity. The present paper deals with this case. We investigate the so-called teacher-student scenario using the replica method by drawing on previous studies~\cite{bachschmid2017statistical,bachschmid2015learning,berg2017statistical}, but refine the theoretical treatment in~\cite{bachschmid2017statistical} for dealing with the teacher with sparse connections. The cavity method is used but the cavity field is assumed to consist of separately treatable signal and noise; the functional form of the associated probability distribution is hypothesized. The network structure of couplings is assumed to be locally tree-like, and our theoretical result is expected to be exact on those networks in the thermodynamic limit. To check the accuracy, numerical experiments are also conducted.

The remainder of this paper is organized as follows. In \Rsec{Formulation}, we review the inverse Ising framework and the statistical mechanical formulation to analyze inference accuracy. We also briefly review the analysis of the fully-connected case given in~\cite{bachschmid2017statistical} to demonstrate how the formulation is applied. In \Rsec{Details}, we extend the method in the statistical mechanical formulation to the sparsely-connected case. Some theoretical treatments developed in~\cite{bachschmid2017statistical} are refined and our ansatz to deal with the teacher model with sparse connections is stated. In \Rsec{Numerical}, our numerical analysis results are provided and compared with our theoretical results to check their accuracy. The final section presents a summary and discussion. 

\section{Formulation} \Lsec{Formulation}
In this section, we briefly review the inverse Ising framework and two associated inference methods: the maximum likelihood (ML) and pseudolikelihood (PL) methods. The class of {\it local learning} is also introduced and explained. In addition, the general framework of the statistical-mechanical analysis proposed in~\cite{bachschmid2017statistical} is presented. We emphasize its technically important points and review the case of the fully-connected Ising model~\cite{bachschmid2017statistical} to demonstrate its application.

\subsection{Inference framework} \Lsec{Inference framework}
Let us consider an Ising model consisting of $N$ spin variables $\V{s}=\lb s_{i}=\pm 1 \rb_{i=1}^{N}$ and obeying the following distribution: 
\be
P_{\rm Ising}\lb \V{s}\big| J,\V{H}\rb=\frac{1}{Z_{\rm Ising}}e^{\sum_{i<j}^{N}J_{ij}s_is_j+\sum_{i=1}^{N}H_is_i},
\ee
where $J\in \mR^{N\times N}$ and $\V{H}\in \mR^{N}$ are termed couplings and external fields, respectively. The setting of the inverse Ising problem is aimed at inferring the couplings and external fields from a given dataset of spin snapshots (samples) $D^{M}\equiv \lbb \V{s}^{(\mu)}\rbb_{\mu=1}^{M}$, where $M$ denotes the dataset size. The representative statistical solution or estimator is the ML estimator defined by
\be
\lbb 
\hat{J}^{\rm ML}(D^M),\hat{\V{H}}^{\rm ML}(D^M)
\rbb
=
\argmin_{J,\V{H}}
\lbb 
-
\sum_{\mu=1}^{M} \log P_{\rm Ising}\lb \V{s}^{(\mu)} \big| J,\V{H}\rb
\rbb.
\Leq{ML}
\ee
Hereafter, the symbol $\hat{\cdot}$ is used to represent an estimator. This canonical estimator has some desirable properties such as {\it consistency}, {\it unbiasedness}, and {\it efficiency} when the dataset size $M$ is large enough while $N$ is kept finite (the asymptotic limit)~\cite{cover2012elements,nguyen2017inverse}. Consistency means that the estimator converges in probability to the true parameter in the large $M$ limit; unbiasedness means that the mean of the estimator (which is a random variable when the dataset is randomly generated) becomes identical to the true parameter; unbiased estimators yielding the lowest mean-squared error are termed efficient. Note that the ML estimator enjoys the unbiasedness and the efficiency only in the asymptotic sense, and for finite size datasets it can have bias and be not efficient.    

In this sense, the ML estimator is good, but is not always appropriate for the inverse Ising framework for some computational reasons. The PL method~\cite{besag1974spatial} overcomes this problem by replacing the likelihood with the conditional distribution $P\lb s_i \big| \V{s}_{\bs i},\V{J}_i,H_i\rb$ for each $s_i$, where $\V{J}_{i}=(J_{ij})_j$ is the coupling vector connected to the $i$th spin, and $\V{s}_{\bs i}$ denotes the spin vector without the $i$th component. The explicit form is 
\be
&&
P\lb s_i \big| \V{s}_{\bs i},\V{J}_i,H_i\rb
=\frac{1}{Z_i}e^{s_{i}\lb \sum_{j(\neq i)}J_{ij}s_j+H_i \rb},
\\ &&
Z_{i}=2\cosh \lb \sum_{j(\neq i)}J_{ij}s_j+H_i \rb.
\ee
The PL estimator is obtained separately for each $i$ by
\be
&&
\lbb 
\hat{\V{J}}^{\rm PL}_i(D^M),\hat{H}_i^{\rm PL}(D^M)
\rbb
=
\argmin_{\V{J}_i,H_i}
\lbb 
-
\sum_{\mu=1}^{M} 
\log P\lb s^{(\mu)}_i \big|\V{s}^{(\mu)}_{\bs i}, \V{J}_i,H_i\rb
\rbb
\Leq{estimator_PL}
\no \\ &&
=
\argmin_{\V{J}_i,H_i}
\lbb 
\sum_{\mu=1}^{M} 
\ell^{\rm PL}\lb s^{(\mu)}_ih_i(\V{s}^{(\mu)}_{\bs i},\V{J}_i,H_i) \rb
\rbb,
\ee
where 
\be
&&
h_i(\V{s}_{\bs i},\V{J}_i,H_i)=\sum_{j(\neq i)}J_{ij}s_j+H_i,
\Leq{effective field}
\\ 
&& 
\ell^{\rm PL}\lb x \rb=-x+\log 2\cosh x.
\Leq{ell_PL}
\ee
Two remarkable properties are held by the PL estimator. The first is its consistency. When the dataset size $M$ is sufficiently large, the PL estimator converges to the true values $\{\hat{\V{J}}^{\rm PL}_i,\hat{H}^{\rm PL}_i\}\to \{\V{J}_i,H_i\}$. The second is its locality. Owing to the factorized nature of PL, each coupling vector $\V{J}_i$ can be assessed independently with low (polynomial) computational cost, which is in contrast to the ML estimator for which the exponentially large computational cost is necessary when assessing the partition function. However, coupling symmetry $J_{ij}=J_{ji}$ is inevitably lost ($\hat{J}^{\rm PL}_{ij}\neq \hat{J}^{\rm PL}_{ji}$ in general).  

This local property of PL also simplifies the theoretical treatment, which inspired explorations of the ``optimal cost function'' in a class of models with the same locality as the PL case~\cite{bachschmid2017statistical,berg2017statistical}. On the basis of these studies, we treat this {\it local learning} class and assume that the cost function argument is the only product of the local spin and effective field. The cost function and corresponding estimator are denoted as $\ell=\ell(s_ih(\V{s}_{\bs i}) )$ and $\lbb \hat{\V{J}}_i,\hat{H}_i \rbb$, respectively. When specifying a certain model in the class, the appropriate superscript will be attached as \Reqs{estimator_PL}{ell_PL}.

We have an additional remark on the lost symmetry in the PL estimator. This limitation of the PL estimator can be recovered by adding PL cost functions of different sites $i$ and $j$ and by performing the minimization with respect to a symmetric variable $J_{ij}$ shared in both terms~\cite{decelle2014pseudolikelihood}. Although our theoretical analysis presented in this paper can also be generalized to this case with a slight increase in the respective computational steps shown in \Rsec{Details}, we restrict ourselves on the simplest case without such a symmetrizing operation. 

\subsubsection{Teacher-student scenario} \Lsec{Teacher-student scenario}
Here, we consider the inverse Ising problem in the teacher-student scenario. Specifically, the dataset $D^{M}\equiv \lbb \V{s}^{(\mu)}\rbb_{\mu=1}^{M}$ is assumed to be independently identically distributed (i.i.d.) samples from a teacher Ising model with the couplings $J^*$ and external fields $\V{H}^*$, and a student Ising model attempts to infer the teacher couplings and fields from the dataset. The inference accuracy is quantified by the residual sum of squares (RSS) between the teacher couplings and student's estimator
\be
\RSS=||\V{J}_i^*-\hat{\V{J}}_i||_2^2=\sum_{j(\neq i)}\lb J_{ij}^*-\hat{J}_{ij} \rb^2
=R^*-2\rho+R,
\Leq{RSS}
\ee
where we defined the following three macroscopic parameters:
\subbe
\Leq{RSSparameters}
\be
&&
R^*=\sum_{j(\neq i)}\lb J_{ij}^* \rb^2,
\\ &&
R=\sum_{j(\neq i)}\lb \hat{J}_{ij} \rb^2,
\\ &&
\rho=\sum_{j(\neq i)} J_{ij}^*\hat{J}_{ij}.
\ee
\subee
These are later associated with the order parameters in the statistical mechanical analysis. 

\subsection{Statistical mechanical analysis} \Lsec{Statistical mechanical}
Here, we explain the statistical mechanical formulation developed in~\cite{bachschmid2017statistical} to analyze the theoretical performance of local learning models. For simplicity of the theoretical analysis, we assume the absence of external fields $\V{H}=\V{0}$ both in the teacher and student models. 

\subsubsection{General framework} \Lsec{General framework}
The basic idea of statistical mechanical analysis is to introduce the following Hamiltonian and Boltzmann distribution induced by the cost function $\ell$:
\be
&&
\mc{H}(\V{J}|\V{D}^{M})=\sum_{\mu=1}^{M}\ell\lb s^{(\mu)}_0h(\V{s}^{(\mu)}_{\bs 0},\V{J}) \rb,
\\ &&
P(\V{J}|\V{D}^{M})=\frac{1}{Z}e^{-\IT \mc{H}(\V{J}|\V{D}^{M})},
\Leq{BMdist}
\ee
where 
\be
&&
h(\V{s}_{\bs 0},\V{J})=\sum_{j=1}^{N-1}J_{j}s_{j},
\\ &&
Z=\Tr{\V{J}}e^{-\IT \mc{H}(\V{J}|\V{D}^{M})}.
\ee
Here, we reorder the spins and focus on the zeroth spin and its coupling vector $\V{J}$. The external field is set to zero as declared above; $\Tr{\V{J}}$ denotes the integration with respect to $\V{J}$ with an appropriate measure, the detailed definition of which is given in \Rsec{Computations}. In the limit $\IT \to \infty$, the Boltzmann distribution converges to a pointwise measure on the estimator $\hat{\V{J}}=\argmin_{\V{J}}\lbb \sum_{\mu=1}^{M}\ell\lb s^{(\mu)}_0h(\V{s}^{(\mu)},\V{J}) \rb \rbb$; thus, the Boltzmann distribution enables the capture of the estimator's properties. We are interested in the thermodynamic limit $N\to \infty$ while keeping $\alpha=M/N=O(1)$; hence, the free energy density $f=-(N\IT)^{-1}\log Z$ will show the self-averaging property. Thus, for typical datasets, the free energy density converges to its average in this limit. The averaged free energy density is denoted as
\be
f^{\rm ave}=-(N\IT)^{-1} \lsb \log Z \rsb_{D^M},
\ee
where the square brackets $\lsb \cdot  \rsb_{D^M}$ denote the average over the dataset, which is the average over the teacher Ising model:
\be
\lsb  \cdot \rsb_{D^M}=
\sum_{\V{s}^{(1)},\cdots,\V{s}^{(M)}} 
\lb \cdot \rb
\prod_{\mu=1}^{M}
P_{\rm Ising}(\V{s}^{(\mu)}|\V{J}^*).
\ee
The average of $\log Z$ over $D^M$ is, however, generally difficult to compute. The replica method is a prescription to overcome this difficulty and is symbolized by the following identity:
\be
f^{\rm ave}
=-(N\IT)^{-1} \lsb \log Z \rsb_{D^M}
=\lim_{n\to 0}-(nN\IT)^{-1} \log \lsb Z^n \rsb_{D^M}.
\ee
Assuming that $n$ is a positive integer, we can rewrite $\lsb Z^n \rsb_{D^M}$ as 
\be
&&
\lsb Z^n \rsb_{D^M}
=\Tr{ \{ \V{J}^a \}_{a=1}^{n} } 
\lbb 
\sum_{\V{s}} 
P_{\rm Ising}(\V{s}|\V{J}^*)
e^{-\IT \sum_{a=1}^{n}\ell\lb s_0h(\V{s}_{\bs 0},\V{J}^a) \rb}
\rbb^M.
\Leq{Z^n_1st}
\ee
Rewriting this by introducing variables $\lb h^a=\sum_{j=1}^{N-1}J_{j}^{a} s_{j} \rb_a$ and $h^*=\sum_{j=1}^{N-1}J_{j}^* s_{j}$, which are hereafter called {\it cavity fields}, we get
\be
&&
\hspace{-12mm}
\lsb Z^n \rsb_{D^M}
=
\Tr{ \{ \V{J}^a \}_{a=1}^{n} } 
\Biggl\{
\sum_{\V{s}} 
\int 
dh^* 
\prod_{a=1}^{n}dh^a
\no \\ && \times
\delta\lb h^*-\sum_{j=1}^{N-1}J_{j}^{*} s_{j} \rb
\prod_{a=1}^{n}
\delta\lb h^a-\sum_{j=1}^{N-1}J_{j}^{a} s_{j} \rb
P_{\rm Ising}(\V{s}\big|\V{J}^*)
e^{-\IT \sum_{a=1}^{n}\ell\lb s_0h^a \rb}
\Biggr\}^M
\no \\ &&
\hspace{-12mm}
=
\Tr{ \{ \V{J}^a \}_{a=1}^{n} } 
\Biggl\{
\sum_{s_0} 
\int 
dh^* 
\prod_{a=1}^{n}dh^a
P_{\rm cav}(h^*,\lbb h^a\rbb_{a=1}^n \big| \V{J}^*,\lbb \V{J}^a\rbb_{a=1}^n)
\frac{1}{Z_0}e^{s_0h^*}
e^{-\IT \sum_{a=1}^{n}\ell\lb s_0h^a \rb}
\Biggr\}^M,
\Leq{Z^n_2nd}
\ee
At the last equality, we took the summation over the spins except for $s_0$, yielding the joint distribution $P_{\rm cav}(h^*,\lbb h^a \rbb_{a=1}^n \big| \V{J}^*,\lbb \V{J}^a\rbb_{a=1}^n)$ of the cavity fields. The normalization constant $Z_0$ is defined through its marginal as 
\be 
Z_0=\int dh^* P_{\rm cav}(h^*|\V{J}^*)2\cosh h^*.
\ee
To proceed further with the computation, we need to specify the functional form of the cavity field distribution and take the average over it. This is possible when the teacher is a fully-connected model, as demonstrated in \cite{bachschmid2017statistical}. We review this result below as it contains some steps essential for the sparsely-connected case.

\subsubsection{Revisit the fully-connected case} \Lsec{Revisit the fully-connected}
When the teacher is a fully-connected model, we can use the central limit theorem and assume that the cavity fields are multivariate Gaussian variables with appropriate covariances and means. In \cite{bachschmid2017statistical}, the authors assumed that the teacher system is in the paramagnetic phase and the replica symmetry (RS) holds in both the student and teacher systems. These assumptions imply that the following four order parameters are sufficient to describe the free energy density:
\subbe
\Leq{order parameters-fc}
\be
&&
Q^* \equiv \sum_{i,j}C_{ij}^{\bs 0} J_{i}^*J^*_j,
\\ &&
Q \equiv \sum_{i,j}C_{ij}^{\bs 0}J_{i}^a J_j^{a},
\Leq{Q_def_fc}
\\ &&
q \equiv \sum_{i,j}C_{ij}^{\bs 0}J_{i}^a J_j^{b},~(a\neq b),
\Leq{q_def_fc}
\\ &&
m \equiv \sum_{i,j}C_{ij}^{\bs 0}J_{i}^*J_j^{a},
\ee
\subee
where $C^{\bs 0}$ is the correlation matrix between the spins:
\be
C_{ij}^{\bs 0}
=\Ave{s_i s_j}^{\bs 0}-\Ave{s_i}^{\bs 0}\Ave{s_j}^{\bs 0}=\Ave{s_i s_j}^{\bs 0},
\ee
where $\Ave{\cdots}^{\bs 0}$ denotes the average over the teacher Ising model without the zeroth spin; the last equality is due to the paramagnetic assumption. The covariances of the cavity fields are described by 
\Req{order parameters-fc} as $\Ave{h^a h^b}^{\bs 0}=Q\delta_{ab}+(1-\delta_{ab})q,~\Ave{h^* h^a}^{\bs 0}=m,~\Ave{\lb h^*\rb^2}^{\bs 0}=Q^*$. Upon assuming this, we can rewrite $\lsb Z^n \rsb_{D^M}$ as
\be
&&
\lsb Z^n \rsb_{D^M}
=
\int dQ dq dm~e^{NS\lb C^{\bs 0},\V{J}^*,Q,q,m\rb+M \log L(Q^*,Q,q,m)},
\ee
where 
\be
&&
\hspace{-10mm}
e^{NS\lb C^{\bs 0},\V{J}^*, Q,q,m\rb}
\equiv
\Tr{ \{ \V{J}^a \}_{a=1}^{n} } 
\prod_{a=1}^{n}
\lbb
\delta\lb
Q - \sum_{i,j}C_{ij}^{\bs 0}J_{i}^a J_j^{a}
 \rb
\delta\lb
m - \sum_{i,j}C_{ij}^{\bs 0}J_{i}^*J_j^{a}
\rb
\rbb
\no \\ && 
\times 
\prod_{a<b}
\delta\lb
q - \sum_{i,j}C_{ij}^{\bs 0}J_{i}^a J_j^{b}
\rb,
\Leq{S_def}
\\ &&
\hspace{-10mm}
L(Q^*,Q,q,m)
\equiv
\sum_{s_0} 
\int 
dh^* 
\prod_{a=1}^{n}dh^a
P_{\rm cav}(h^*,\lbb h^a \rbb_{a=1}^n \big| Q^*,Q,q,m)
\frac{1}{Z_0}e^{s_0h^*}
e^{-\IT \sum_{a=1}^{n}\ell\lb s_0h^a \rb}.
\Leq{L_def}
\ee
Deferring the detailed computations to \Rsec{Computations}, we immediately have the result in the limit $n\to 0$:
\be
&&
\hspace{-1cm}
\lim_{n\to 0}\frac{1}{n}S\lb C^{\bs 0},\V{J}^*, Q,q,m\rb
=
\frac{1}{2}\lbb 
\frac{Q-m^2/Q^*}{Q-q}+\log 2\pi +\log (Q-q)
-\frac{1}{N}\Tr{}\log C^{\bs 0}
\rbb,
\Leq{S_fc_nzero}
\\ &&
\hspace{-1cm}\lim_{n\to 0}\frac{1}{n}\log L\lb Q^*,Q,q,m\rb
=
\int Dz~e^{\sqrt{\frac{m^2}{q}}z-\frac{1}{2}\frac{m^2}{q}}
\log \int Dv~e^{-\IT \ell \lb \sqrt{Q-q}v+\sqrt{q}z \rb},
\Leq{L_fc_nzero}
\ee
where we introduce a Gaussian measure
\be
\int Dx(\cdots) \equiv \int_{-\infty}^{\infty}\frac{dx}{\sqrt{2\pi}}~e^{-\frac{1}{2}x^2}(\cdots).
\ee
Further, we take the limit $\IT \to \infty$, which requires the following relation:
\be
\lim_{\IT \to \infty}\IT(Q-q)=\chi=O(1).
\Leq{chi}
\ee
After straightforward calculations, we get
\be
&&
f^{\rm ave}\lb \IT \to \infty \rb
=
\lim_{\IT \to \infty}
-\frac{1}{\IT}
\lb 
\lim_{n\to 0}\frac{1}{n}S\lb C^{\bs 0},\V{J}^*, Q,q,m\rb
+
\alpha\lim_{n\to 0}\frac{1}{n}\log L\lb Q^*,Q,q,m\rb
\rb
\no \\ &&
=
-
\Extr{Q,\chi,m}
\lbb
\frac{1}{2}\frac{Q-m^2/Q^*}{\chi}+\alpha \int Dz \max_{y}\lb -\frac{\lb y-\sqrt{Q}z-m \rb^2}{2\chi}-\ell(y) \rb
\rbb,
\ee
where $\alpha=M/N$, and $\Extr{x}$ denotes the extremization condition with respect to $x$. The extremization condition yields the following equations of state (EOS):
\subbe
\Leq{EOS_fc}
\be
&&
0=\frac{1}{\chi}-\frac{\alpha}{\sqrt{Q}}\int Dz z \Part{\ell}{y}{}\Bigg|_{y=\hat{y}},
\\ &&
0=-\frac{m}{Q^*\chi}-\alpha \int Dz \Part{\ell}{y}{}\Bigg|_{y=\hat{y}},
\\ &&
0=-\frac{1}{\chi^2}\lb Q-\frac{m^2}{Q^*} \rb+ \alpha \int Dz  \lb \Part{\ell}{y}{}\Bigg|_{y=\hat{y}} \rb^2,
\ee
\subee
where
\be
\hat{y}(z,Q,\chi,m)=\argmax_{y}\lb -\frac{\lb y-\sqrt{Q}z-m \rb^2}{2\chi}-\ell(y) \rb.
\ee
Using the solution of \Req{EOS_fc}, we can evaluate the macroscopic parameters \NReq{RSSparameters} by
\subbe
\Leq{R and rho}
\be
&&
R=\lb Q-\frac{m^2}{Q^*} \rb \frac{1}{N}\Tr{}\lb C^{\bs 0} \rb^{-1}+R^*\lb \frac{m}{Q^* }\rb^2,
\\ &&
\rho=R^*\frac{m}{Q^*}.
\ee
\subee
These relations can be derived by a standard technique using auxiliary variables~\cite{bachschmid2017statistical} and the derivation is given in \Rsec{Derivation}. Once the values of $R$ and $\rho$ are obtained, the RSS is eventually computed by \Req{RSS}. Note that the values of $Q^*$ and $R^*$ are directly determined by the problem setting, as well as the inverse correlation function $\lb C^{\bs 0}\rb^{-1}$. To obtain these quantities, we need to separately solve the direct problem.

\section{Details of the sparsely-connected case} \Lsec{Details}
This section provides the extension of the above result to the sparsely-connected case, which is the main contribution of this paper. To this end, we introduce an ansatz about the estimator's behavior as well as the functional form of the cavity field distribution. Under the ansatz, the cavity field is decomposed into a signal and a noise, and it is shown that the noise part obeys essentially the same EOS as the fully-connected case. To complete the computation under the ansatz, the tree-like structure of the coupling network of the teacher model is employed.

\subsection{Ansatz for the sparse case} \Lsec{Ansatz for}
In contrast to the fully-connected case, the cavity field distribution in the sparse case cannot be regarded as Gaussian; the distribution of $h^*$ actually becomes the sum of a few pointwise measures, which is far from Gaussian. Hence, we need a new ansatz to handle the cavity field distribution in the sparse case. 

To obtain an idea of how to resolve this, let us consider an ideal situation where we know which couplings are nonzero. Let us suppose that the zeroth spin is connected to $c=O(1)$ neighboring spins, and introduce two sets of indices $\Omega=\{i|J^*_i \neq 0,~i\in \{1,\cdots,N-1\} \}$ and  $\bar{\Omega}=\{i|J^*_i = 0,~i\in \{1,\cdots,N-1\} \}$, where $\Omega~(\bar{\Omega})$ is called the active (inactive) set; $|\Omega|=c$ and $|\bar{\Omega}|=N-1-c$. If we know $\Omega$ and $\bar{\Omega}$ in advance, then the inference should be conducted only on $\{J_{i}|i\in \Omega \}$. Accordingly, the number of variables to be inferred is just $c=O(1)$; hence, the dataset size $M=O(N)$ is sufficiently large. Thus, we can apply the asymptotic theory of statistics, which implies that an estimator in this ideal case behaves as
\be
\hat{J}^{\rm oracle}_i=
\left\{
\begin{array}{cc}
J_{i}^*+\Delta_{i}  & (i \in \Omega)    \\
0  & (i \in \bar{\Omega})  
\end{array}
\right.,
\Leq{oracle}
\ee
this is called an {\it oracle} estimator. The ``error'' from the true solution, $\Delta_{i}$, is a random variable. In the local learning class with appropriate ({\it consistent}) cost functions such as PL~\cite{hyvarinen2006consistency}, $\Delta_i$ is considered to be zero mean with variance decreasing at the rate of $O(M^{-1})=O(N^{-1})$. The RSS is written as $\RSS=\sum_{i \in \Omega}\Delta_{i}^2=O(N^{-1})$, and vanishes in the thermodynamic limit.

Based on these observations about the oracle estimator, we assume that the (non-oracle) estimator obtained from consistent cost functions obeys the following form: 
\be
\hat{J}_i\doteq
\left\{
\begin{array}{cc}
\bar{J}_{i}+\Delta_{i}  & (i \in \Omega)    \\
\Delta_{i}  & (i \in \bar{\Omega})  
\end{array}
\right.,
\Leq{key ansatz1}
\ee
where we again assume that $\Delta_i$ is a random variable which is asymptotically zero mean with variance scaled as $O(N^{-1})$; the correlations among $\{\Delta_i\}_i$ are also assumed to be sufficiently weak. The quantity $\bar{J}_{i}$ is interpreted as the mean value of the estimator and will deviate from the true value $J_{i}^*$ owing to the extensive number of noise terms $\{\Delta_i\}_i$. The values of $\{ \bar{J}_{i}\}_{i\in \Omega}$ are later computed by taking the minimization condition of the free energy as the order parameters. The applicable range of this ansatz is discussed in \Rsec{Applicable range}.

Let us examine the consequence of the ansatz. The RSS can be written as
\be
\RSS
\approx 
\sum_{i \in \Omega}\lb J_i^*-\bar{J}_i \rb^2
+
\sum_{i \in \bar{\Omega}}\Delta_i^2.
\Leq{RSS_sparse}
\ee
Now, there are two non-negligible contributions to the RSS coming from the bias in $\Omega$ and the noise in $\bar{\Omega}$; the RSS remains finite even in the limit $N\to \infty$ in contrast to the ideal case. The ansatz also allows us to decompose the cavity field as
\be
&&
h^a=h_{\Omega}+h^a_{\Delta},
\Leq{h^a_decomposition}
\\ &&
h_{\Omega}\equiv \sum_{j\in \Omega}\bar{J}_{j} s_j,~~
h^a_{\Delta}\equiv\sum_{j}\Delta^a_j s_j 
\approx \sum_{j\in \bar{\Omega}} \Delta^a_j s_j,
\Leq{h^a_noise}
\ee
where $h^a_{\Delta}$ is termed as the ``noise'' part. Furthermore, we can assume that $h_{\Omega}$ and $h^a_{\Delta}$ are asymptotically independent in the limit $N\to \infty$. This assumption is reasonable, and a schematic to explain this is given in \Rfig{independency}.
\begin{figure}[htbp]
\begin{center}
\includegraphics[height=0.35\columnwidth,width=0.485\columnwidth]{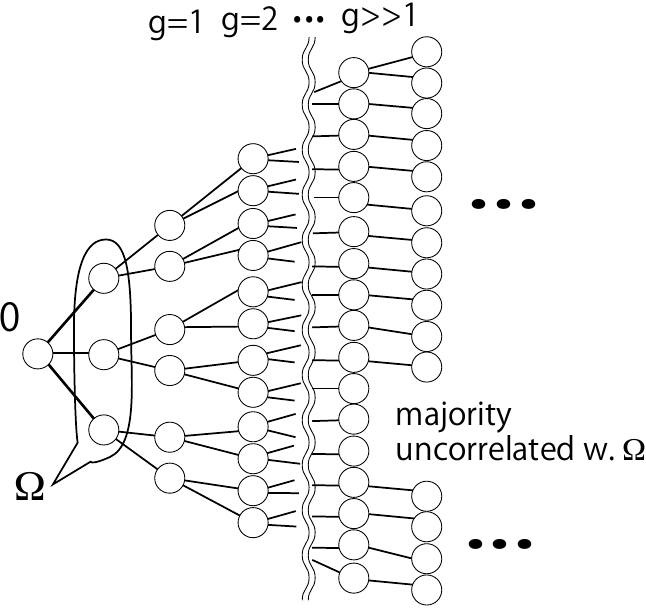}
\caption{Schematic explaining the independence between $(s_0,\V{s}_{\Omega})$ and $h^{a}_{\Delta}$. The majority in the sum in $h_{\Delta}^a$ is uncorrelated with spins in $\Omega$ and dominates $h^a_{\Delta}$ in the thermodynamic limit.}
\Lfig{independency}
\end{center}
\end{figure}
Owing to the tree-like structure, we can define the generation $g$ of a spin $s$ from $\Omega$ as the shortest path length between $s$ and any spin in $\Omega$ along the network. As $g$ grows, the correlation with $\{s_i | i \in \Omega\}$ decays exponentially fast, while the number of spins belonging to generation $g$ exponentially increases. If the correlation decay is sufficiently faster than the increase of the spins, then the majority of spins in the network can be regarded as uncorrelated with $\Omega$. Some terms in $h^{a}_{\Delta}$ are certainly correlated with $h^*$, but their contribution is $O(\sqrt{N}^{-1})$ because $\Delta_i=O(\sqrt{N}^{-1})$ and the number of correlating terms is $O(1)$ owing to the sufficiently fast decay of correlations. Hence, the contribution of correlating terms vanishes and the uncorrelated majority with $\Omega$ completely dominates $h^{a}_{\Delta}$ in the thermodynamic limit. These observations indicate that \Req{Z^n_1st} can be now decomposed as follows:
\be
\hspace{-12mm}
&&
\lsb Z^n \rsb_{D^M}
=
\Tr{ \{ \V{J}^a \}_{a=1}^{n} } 
\Biggl\{
\sum_{\V{s}} 
\int 
P_{\rm Ising}(\V{s}\big|\V{J}^*)
e^{-\IT \sum_{a=1}^{n}\ell\lb s_0h^a \rb}
\Biggr\}^M
\no \\ &&
\approx
\Tr{ \{ \V{\Delta}^a \}_{a=1}^{n} } 
\Biggl\{
\sum_{\V{s}} 
\int \prod_{a=1}^{n} dh^{a}_{\Delta}
P_{\rm Ising}(\V{s}|\V{J}^*)
\delta\lb h^a_{\Delta}-\sum_{j\in \bar{\Omega}}\Delta_{j}^{a} s_{j} \rb
e^{-\IT \sum_{a=1}^{n}\ell\lb s_0\lb \sum_{j\in \Omega}\bar{J}_j s_j+h_{\Delta}^a \rb \rb}
\Biggr\}^M
\no \\ &&
=
\Tr{ \{ \V{\Delta}^a \}_{a=1}^{n} } 
\Biggl\{
\sum_{s_0,\V{s}_{\Omega}} 
\int \prod_{a=1}^{n} dh^{a}_{\Delta}
P(s_{0},\V{s}_{\Omega},\{h^a_{\Delta}\}_a|\V{J}^*,\{\V{\Delta}^a \}_a)
e^{-\IT \sum_{a=1}^{n}\ell\lb s_0\lb \sum_{ j \in \Omega } \bar{J}_j s_j+h_{\Delta}^a \rb \rb}
\Biggr\}^M
\no \\ &&
\approx
\Tr{ \{ \V{\Delta}^a \}_{a=1}^{n} } 
\Biggl\{
\sum_{s_0,\V{s}_{\Omega}} 
\int \prod_{a=1}^{n} dh^{a}_{\Delta}
P_{\rm Ising}(s_{0},\V{s}_{\Omega}|\V{J}^*)P_{\rm cav}(\{h^a_{\Delta}\}_a|\{\V{\Delta}^a\}_a)
\no \\ &&
\times e^{-\IT \sum_{a=1}^{n}\ell\lb s_0\lb \sum_{ j \in \Omega } \bar{J}_j s_j+h_{\Delta}^a \rb \rb}
\Biggr\}^M.
\Leq{Z^n_sparse}
\ee
In the second line, we performed the variable transformation $\V{\Delta}^a=\V{J}^a-\bar{\V{J}}$ and neglected the contribution $\sum_{j\in \Omega}\Delta_j^a s_{j}$ in $h^{a}_{\Delta}$ as \Req{h^a_noise}. In the third line, we denoted $\V{s}_{\Omega}=\{s_i|i\in \Omega\}$ and $\V{s}_{\bar{\Omega}}=\{s_i| i \in \bar{\Omega}\}$, and performed the sum over $\V{s}_{\bar{\Omega}}$, yielding the joint distribution $P(s_{0},\V{s}_{\Omega},\{h^a_{\Delta}\}_a|\V{J}^*,\{\V{\Delta}^a \}_a)$. In the fourth line, we used the asymptotic uncorrelatedness between $h_{\Delta}^a$ and $(s_0,\V{s}_{\Omega})$ discussed so far. 

Now, the central limit theorem can be applied to the noise part $\{h^a_{\Delta}\}_a$, and they can be regarded as Gaussian variables. As the fully-connected case, two order parameters describing their covariances are introduced:
\be
&&
Q \equiv \sum_{i,j}C_{ij}^{\bs 0}\Delta_{i}^a \Delta_j^{a},
\\ &&
q \equiv \sum_{i,j}C_{ij}^{\bs 0}\Delta_{i}^a \Delta_j^{b},~(a\neq b).
\ee
Counterparts of $m$ and $Q^*$ are unnecessary because the dependence on $(s_0,\V{s}_{\Omega})$ is separately and explicitly treated in the present formulation. Then,
\be
&&
\lsb Z^n \rsb_{D^M}
\approx
\int dQ dq ~e^{NS\lb C^{\bs 0},Q,q\rb+M \log L(\V{J}^*,\bar{\V{J}},Q,q)},
\ee
where 
\be
&&
\hspace{-15mm}
e^{NS\lb C^{\bs 0}, Q,q\rb}
\equiv
\Tr{ \{ \V{\Delta}^a \}_{a=1}^{n} } 
\prod_{a=1}^{n}
\delta\lb
Q - \sum_{i,j}C_{ij}^{\bs 0}\Delta_{i}^a \Delta_j^{a}
 \rb
\prod_{a<b}
\delta\lb
q - \sum_{i,j}C_{ij}^{\bs 0}\Delta_{i}^a \Delta_j^{b}
\rb,
\Leq{S_def_sparse}
\\ &&
\hspace{-15mm}
L(\V{J}^*,\bar{\V{J}},Q,q)=
\sum_{s_0,\V{s}_{\Omega}} 
\int 
\prod_{a=1}^{n}dh^a_{\Delta}
P_{\rm Ising}(s_{0},\V{s}_{\Omega}|\V{J}^*)
P_{\rm cav}( \lbb h^a_{\Delta} \rbb_{a=1}^n \big| Q,q)
\no \\ &&
\times 
e^{-\IT \sum_{a=1}^{n}\ell\lb s_0\lb \sum_{j\in \Omega}\bar{J}_{j}s_j+h^a_{\Delta}  \rb \rb}.
\Leq{L_def_sparse}
\ee
Again, using the techniques in~\Rsec{Computations} we get 
\be
&&
\hspace{-1.5cm}
\lim_{n\to 0}\frac{1}{n}S\lb C^{\bs 0}, Q,q\rb
=
\frac{1}{2}\lbb 
\frac{Q}{Q-q}+\log 2\pi +\log(Q-q)
-\frac{1}{N}\Tr{}\log C^{\bs 0}
\rbb,
\Leq{S_sparse_nzero}
\\ &&
\hspace{-1.5cm}\lim_{n\to 0}\frac{1}{n}\log L\lb \V{J}^*,\bar{\V{J}},Q,q\rb
=
\sum_{s_0,\V{s}_{\Omega}}
P_{\rm Ising}(s_0,\V{s}_{\Omega}|\V{J}^*)
\no \\ &&
\times
\int Dz \log \int Dv~e^{-\IT \ell \lb s_0\lb \sum_{j \in \Omega}\bar{J}_{j}s_j+\sqrt{Q-q}v+\sqrt{q}z \rb \rb}.
\Leq{L_sparse_nzero}
\ee
Employing the relation \NReq{chi} and taking the $\IT \to \infty $ limit, we get
\be
&&
f^{\rm ave}
=
-\Extr{Q,\chi}
\Bigg\{
\frac{1}{2}\frac{Q}{\chi}
\no \\ &&
+
\alpha 
\sum_{s_0,\V{s}_{\Omega}}
P_{\rm Ising}(s_0,\V{s}_{\Omega}|\V{J}^*)
\int Dz
\max_{y}
\lb 
-\frac{\lb y-s_0(\sqrt{Q}z+\sum_{j \in \Omega}\bar{J}_{j}s_j) \rb^2}{2\chi}
- \ell \lb y \rb 
\rb
\Bigg\}.
\Leq{f_sparse}
\ee
The extremization condition with respect to $Q$ and $\chi$ gives
\subbe
\Leq{EOS_sparse}
\be
&&
0=\frac{1}{\chi}-\frac{\alpha}{\sqrt{Q}}
\sum_{s_0,\V{s}_{\Omega}}
P_{\rm Ising}(s_0,\V{s}_{\Omega}|\V{J}^*)
s_0
\int Dz
 z \Part{\ell}{y}{}\Bigg|_{y=\hat{y}},
\\ &&
0=-\frac{Q}{\chi^2}+ \alpha 
\sum_{s_0,\V{s}_{\Omega}}
P_{\rm Ising}(s_0,\V{s}_{\Omega}|\V{J}^*)
\int Dz
  \lb \Part{\ell}{y}{}\Bigg|_{y=\hat{y}} \rb^2,
\ee
\subee
where
\be
\hat{y}(z,s_0,\V{s}_{\Omega}|Q,\chi,\{\bar{J}_j\}_{j\in \Omega})=\argmax_{y}\lb -\frac{\lb y-s_0\lb \sqrt{Q}z+\sum_{j \in \Omega}\bar{J}_{j}s_j \rb \rb^2}{2\chi}-\ell(y) \rb.
\Leq{yhat}
\ee
Further, the mean estimates $\{\bar{J}_{j}\}_{j\in \Omega}$ are also evaluated by the extremization condition. The result for $\bar{J}_{j}$ is given by
\be
&&
0=
\sum_{s_0,\V{s}_{\Omega}}P_{\rm Ising}(s_0,\V{s}_{\Omega}|\V{J}^*)
\int Dz \Part{\ell}{y}{}\Bigg|_{y=\hat{y}} 
s_0s_j
\Leq{Jbar}.
\ee
Using the parameters $Q,\chi,\{\bar{J}_i\}_{i\in\Omega}$ computed by \Reqss{EOS_sparse}{Jbar}, we can evaluate the RSS which is expressed in the present setting as
\be
\RSS
\approx \sum_{i\in \Omega}(J_{i}^*-\bar{J}_{i})^2+\sum_{i\in \bar{\Omega}}{\Delta}_i^2
=
\sum_{i\in \Omega}(J_{i}^*-\bar{J}_{i})^2+\frac{Q}{N}\Tr{}\lb C^{\bs 0} \rb^{-1}.
\Leq{RSS_sparse2}
\ee
The quantity $\lb C^{\bs 0} \rb^{-1}$ will be computed in another discussion on the direct problem as the fully-connected case. In addition, $P_{\rm Ising}(s_0,\V{s}_{\Omega}|\V{J}^*)$ will also be assessed separately in the sparse case. These points are addressed in the next subsection. 

\subsection{Direct problem's properties} \Lsec{Direct problem's} 
The inverse problem essentially requires certain information from its direct problem counterpart. Necessary information to compute the quantities of interest depends on the system's properties. In the fully-connected case, two-body quantities such as $\lb C^{\bs 0}\rb^{-1}$ and $\sum_{i,j}C^{\bs 0}_{ij}J_{i}^*J_{j}^*$ are sufficient. However, in the sparse case, higher-order information is needed because the central limit theorem does not fully dominate the system. Hence, the functional form of $P_{\rm Ising}(s_0,\V{s}_{\Omega})$ becomes necessary, as seen in \Req{f_sparse}. Techniques for computing such quantities in the sparse case largely advanced in the '90--'00s. Here, we quote a portion of the results to compute the necessary quantities. For readers interested in the detailed techniques, please refer to~\cite{opper2001advanced,mezard2009information}. Although these techniques are applied in general situations, to obtain compact analytic forms of the quantities of interest, we rely on the assumptions that the teacher model is in the paramagnetic phase and the external fields are absent.

\subsubsection{Marginal distribution of the teacher model} \Lsec{Marginal distribution on} 
The marginal distribution $P_{\rm Ising}(s_0,\V{s}_{\Omega}|\V{J}^*)$ is computed by marginalizing the whole distribution $P_{\rm Ising}(\V{s}|\V{J}^*)$ with respect to $\V{s}_{\bar{\Omega}}$. In general, this operation requires nontrivial computations and the resultant distribution becomes dependent on parameters among the marginalized spins. However, under the present assumptions, such dependencies do not exist and the expression becomes rather simple: 
\be
P_{\rm Ising}(s_0,\V{s}_{\Omega}|\V{J}^*)
=\frac{1}{Z_{\Omega_c}}e^{s_{0}\sum_{j\in\Omega}J_{j}^*s_j }
,~
Z_{\Omega_c}=\sum_{s_0,\V{s}_{\Omega}}e^{s_{0}\sum_{j\in\Omega}J_{j}^*s_j },
\Leq{cavdist_sparse}
\ee
where $\Omega_{c}$ denotes the union index set of $0$ and $\Omega$. This form is applied to \Reqs{EOS_sparse}{Jbar} to obtain the order parameters.

\subsubsection{Inverse correlation function} \Lsec{Inverse correlation function} 
Next, we compute the inverse correlation function $C^{-1}$; hereafter, we treat the whole system and discard the superscript $^{\bs 0}$ as it is not essential. 

On the tree-like networks, the so-called {\it Bethe approximation} gives an exact result as the system size tends to infinity, and it provide a compact analytic form of the inverse correlation function. The derivation is given in~\cite{ricci2012bethe,nguyen2012bethe} and we here just quote the result. Under the present assumptions without magnetizations and external fields, it becomes
\be
\lb C^{-1} \rb_{ij}
=\lb \sum_{k \in \partial i }\frac{1}{1-\tanh^2(J_{ik})}-(c_i-1) \rb \delta_{ij}
-\frac{\tanh(J_{ij})}{1-\tanh^2(J_{ij})}(1-\delta_{ij}),
\Leq{C_inverse}
\ee
where $\partial i$ denotes the index set of nodes connected to $i$ and $c_i$ denotes the connectivity or the number of edges connecting to node $i$. 

\subsection{Applicable range of the ansatz } \Lsec{Applicable range} 
Here we consider the applicable range of the ansatz \NReq{key ansatz1}. This ansatz is a strong statement since it allows us to not think of any possible biases of the estimators outside the active set $\Omega$. When is this valid? How does it relate to the tree-like network structure? 

To obtain answers to these questions, we rethink \Req{Jbar}. An important observation is that this equation is merely the zero-gradient condition of $\ell$ with respect to $J_{j},(j\in \Omega)$ averaged over the datasets, as shown below. Denoting the empirical average on the dataset $D^M$ by $\overline{\cdots}^{D^M}$, and using the statistical mechanical analysis explained so far, we can write the zero-gradient condition with respect to $J_{j}$ for $j\in \Omega$ as
\be
&&
0=
\overline{\Part{ \ell( s_{0}h( \V{s}_{\bs 0},\V{J}) ) }{J_j}{}\Biggr|_{\V{J}=\hat{\V{J}}(D^M) }}^{D^M}
=
\overline{ \Part{ \ell( y) }{y}{}s_0s_j }^{D^M},
\ee
where we put $y=s_{0}h( \V{s}_{\bs 0},\hat{\V{J}})=s_0\lb  \sum_{i}\hat{J}_i s_i \rb$. With this expression, we replace the estimator $\hat{\V{J}}$ with the average over \Req{BMdist}, take the average over the dataset, and use the replica method. The result is 
\be
&&
0=
\lim_{\beta \to \infty}
\lim_{n\to 0}
\Tr{\{ \V{J}^a \}_{a=1}^{n}}
\lb \sum_{\V{s}}P_{\rm Ising}(\V{s}\big|\V{J}^*)e^{-\beta \sum_{a=1}^{n} \ell(y^a) }\rb^{M-1}
\no \\ && \times
\lb 
\sum_{\V{s}}
P_{\rm Ising}(\V{s}\big|\V{J}^*)
e^{-\beta \sum_{a=1}^{n} \ell(y^a) }
\frac{\partial \ell( y^1) }{\partial y^1}
s_0 s_j
\rb.
\Leq{grad-replica}
\ee
where $y^{a}\equiv s_0\lb \sum_{i}J_i^as_j \rb$. The ansatz \NReq{key ansatz1} and RS used in \Rsec{Ansatz for}, in short, say  
\be
&&
y^a
\overset{\mathrm{ansatz}}{=}
s_0\lbb \sum_{i \in \Omega}\bar{J}_i s_i+h_{\Delta}^a \rbb
\overset{\mathrm{RS}}{=}
s_0\lbb \sum_{ i \in \Omega}\bar{J}_i s_i+\sqrt{Q-q}v^a+\sqrt{q}z \rbb. 
\Leq{ansatz in a nutshell}
\ee
where $v^a,z\sim \mathcal{N}(0,1)$. Applying this form and following the same line of computations as in \Rsec{Ansatz for}, we get
\be
&&
\hspace{-1cm}
\lb 
\sum_{\V{s}}
P_{\rm Ising}(\V{s}\big|\V{J}^*)
e^{-\beta \sum_{a=1}^{n} \ell(y^a) }
\frac{\partial \ell( y^1) }{\partial y^1}
s_0 s_j
\rb
\no \\ &&
\hspace{-1cm}
\overset{N\to \infty}{\longrightarrow}
\sum_{s_0,\V{s}_{\Omega}}
P_{\rm Ising}(s_0,\V{s}_{\Omega} \big|\V{J}^*)
\int Dz
\lb \int Dv
e^{-\beta \ell(y(z,v)) }
\rb^n
\frac{
\int Dv^1 e^{-\beta \ell(y(z,v^1)) }
\frac{\partial \ell( y^1) }{\partial y^1}
s_0 s_j
}
{
\int Dv^1 e^{-\beta \ell(y(z,v^1)) }
}
\no \\ &&
\hspace{-1cm}
\overset{n\to 0}{\longrightarrow}
\sum_{s_0,\V{s}_{\Omega}}
P_{\rm Ising}(s_0,\V{s}_{\Omega} \big|\V{J}^*)
\int Dz
\frac{
\int Dv e^{-\beta \ell(y(z,v)) }
\frac{\partial \ell( y) }{\partial y}
s_0 s_j
}
{
\int Dv e^{-\beta \ell(y(z,v)) }
}
\no \\ &&
\hspace{-1cm}
\overset{\beta \to \infty}{\longrightarrow}
\sum_{s_0,\V{s}_{\Omega}}
P_{\rm Ising}( s_0,\V{s}_{\Omega} \big|\V{J}^* )
\int Dz
\frac{\partial \ell( y) }{\partial y} \Biggr|_{y=\hat{y}}
s_0 s_j,
\ee
and the factor $\lb \sum_{\V{s}}P_{\rm Ising}(\V{s}\big|\V{J}^*)e^{-\beta \sum_{a=1}^{n} \ell(y^a) }\rb^{M-1}$ in \Req{grad-replica} becomes unity when taking the $n\to 0$ limit, yielding the identical result to \Req{Jbar}. 

This computation naturally leads to the following question: Should we compute all the zero-gradient conditions not only for $ \Omega$ but also for $ \bar{\Omega}$? This point is important because if this is the case, then the ansatz \NReq{key ansatz1} is insufficient as it only suffices those for the active set $j\in \Omega$. To be consistent, the answer is considered to be yes in general; hence, we need to take into account the zero-gradient conditions for $k \in \bar{\Omega}$. This implies that the ansatz \NReq{key ansatz1} should be modified and we need to introduce mean estimates $\bar{J}_k$ for $k \in \bar{\Omega}$ in general situations. 

Fortunately, if the network is tree-like, we can show that all the zero-gradient conditions are automatically satisfied once those for $\forall j\in \Omega$ are met. Hence, the ansatz \NReq{key ansatz1} is consistent on such networks; we show proof of this below. For technical reasons, we recover the external field $\V{H}^*$ in the remaining of this subsection. When the external field exists, the student model should also have the external field variable, and hence the replica result is slightly modified. That modification is accomplished by replacing $\lb \sum_{j\in \Omega}\bar{J}_js_j \rb$ with $\lb \sum_{j\in \Omega}\bar{J}_js_j +\bar{H}_0 \rb$ in \Reqs{f_sparse}{yhat} and \NReq{ansatz in a nutshell}. Here, $\bar{H}_0$ denotes the mean estimate of the external field variable acting on the focused spin $s_0$ of the student model, and is determined by the extremization condition of the free energy, yielding
\be
&&
0=
\sum_{s_0,\V{s}_{\Omega}}P_{\rm Ising}(s_0,\V{s}_{\Omega}|\V{J}^*,\V{H}^*)
\int Dz \Part{\ell}{y}{}\Bigg|_{y=\hat{y}} 
s_0.
\Leq{H0}
\ee

Under the above setup, we show the consistency of \Req{key ansatz1} on tree-like networks. The first step is to write down the zero-gradient condition for $k\in \bar{\Omega}$. The result of applying the averages and replica method is the replacement of $s_j$ with $s_k$ in \Req{grad-replica}. With this expression, we perform the following transformation:
\be
&&
\hspace{-1.5cm}
\sum_{\V{s}}
P_{\rm Ising}(\V{s}\big|\V{J}^*,\V{H}^*)
e^{-\beta \sum_{a=1}^{n} \ell(y^a) }
\frac{\partial \ell( y^1) }{\partial y^1}
s_0 s_k
=
\frac{\partial }{\partial H_k^*}
\lb
\sum_{\V{s}}
P_{\rm Ising}(\V{s}\big|\V{J}^*,\V{H}^*)
e^{-\beta \sum_{a=1}^{n} \ell(y^a) }
\frac{\partial \ell( y^1) }{\partial y^1}
s_0 
\rb
\hspace{-1cm}
\no \\ &&
+
\lb
\sum_{\V{s}}
P_{\rm Ising}(\V{s}\big|\V{J}^*,\V{H}^*)
e^{-\beta \sum_{a=1}^{n} \ell(y^a) }
\frac{\partial \ell( y^1) }{\partial y^1}
s_0 
\rb \Ave{s_k},
\ee 
where $\Ave{\cdots}$ denotes the average over $P_{\rm Ising}(\V{s}\big|\V{J}^*,\V{H}^*)$. By following the same computations so far, the second term vanishes in the limit $\lim_{\beta \to \infty}\lim_{n \to 0}\lim_{N \to \infty}$ because the coefficient converges to the right-hand side of \NReq{H0} giving zero. Meanwhile, in the same limit, the first term can be transformed as
\be
&&
\frac{\partial }{\partial H_k^*}
\lb
\sum_{\V{s}}
P_{\rm Ising}(\V{s}\big|\V{J}^*,\V{H}^*)
e^{-\beta \sum_{a=1}^{n} \ell(y^a) }
\frac{\partial \ell( y^1) }{\partial y^1}
s_0 
\rb
\no \\ &&
\to 
\frac{\partial }{\partial H_k^*}
\lb 
\sum_{s_0,\V{s}_{\Omega}}
P_{\rm Ising}( s_0,\V{s}_{\Omega} \big|\V{J}^*,\V{H}^*)
\int Dz
\frac{\partial \ell( y) }{\partial y} \Biggr|_{y=\hat{y}}
s_0
\rb,
\ee
and the dependence on $H_{k}^*$ appears only in the marginal distribution $P_{\rm Ising}( s_0,\V{s}_{\Omega} \big|\V{J}^*,\V{H}^*)$. On tree-like networks, the marginal distribution necessarily takes the following form:
\be
P_{\rm Ising}( s_0,\V{s}_{\Omega} \big|\V{J}^*,\V{H}^*)
=\frac{1}{Z}e^{s_0\lb \sum_{i\in \Omega}J_i^*s_i +H_0^* \rb+\sum_{i\in \Omega}h_{i}^{\bs 0} s_i},
\Leq{marginal-tree}
\ee
where $h_i^{\bs 0}$ is the effective field obtained by marginalizing the descendant spins of $i$, and is usually termed as cavity field. An important point of \Req{marginal-tree} is the absence of higher-order interactions among active set spins because of the tree-like structure. Hence, the differentiation of $H_k^*$ appears only through that of the effective fields. Furthermore, owing to the tree-like structure, only one of the effective fields is dependent on $H^*_k$. Specifying the corresponding index as $j(\in \Omega)$, we get
\be
&&
\frac{\partial }{\partial H_k^*}
\lb 
\sum_{s_0,\V{s}_{\Omega}}
P_{\rm Ising}( s_0,\V{s}_{\Omega} \big|\V{J}^*,\V{H}^*)
\int Dz
\frac{\partial \ell( y) }{\partial y} \Biggr|_{y=\hat{y}}
s_0
\rb
\no \\ &&
=
\frac{\partial h_j^{\bs 0}}{\partial H_k^*}
\frac{\partial }{\partial h_j^{\bs 0}}
\lb 
\sum_{s_0,\V{s}_{\Omega}}
P_{\rm Ising}( s_0,\V{s}_{\Omega} \big|\V{J}^*,\V{H}^*)
\int Dz
\frac{\partial \ell( y) }{\partial y} \Biggr|_{y=\hat{y}}
s_0
\rb
\no \\ &&
=
\frac{\partial h_j^{\bs 0}}{\partial H_k^*}
\lbb
\mathrm{(r.h.s.~of~eq.~\NReq{Jbar})-(r.h.s.~of~eq.~\NReq{H0})}\times \Ave{s_j}
\rbb
=0.
\ee
Hence, the zero-gradient conditions on the inactive set $\bar{\Omega}$ are  satisfied once those of the active set $\Omega$ hold, proving our statement. 

The above proof also provides a perspective for loopy graphs. If loops exist, then higher-order interactions emerge in $P_{\rm Ising}(s_0,\V{s}_{\Omega})$;  they generally depend on $H_{k}^*$ in a complex manner and yield some additional terms as a result of differentiation. In such situations, additional mean estimates $\bar{J}_{k}$ for $k\in \bar{\Omega}$ will be necessary to satisfy the corresponding zero-gradient conditions; however, treating all variables in $\bar{\Omega}$ is clearly infeasible. Tailoring good approximations in such cases may be interesting in future work, although in \Rsec{Square lattice} we show an example in which our present theoretical treatment becomes a good approximation even for loopy graphs.

\section{Numerical experiments} \Lsec{Numerical}
In this section, we conduct numerical experiments to check the accuracy of the theoretical computations. The actual behavior of the order parameters and related quantities depends on the details of the coupling ensembles. Hence,  we treat the regular random (RR) graph and Erd\H{o}s--R\'enyi (ER) graph as representative examples of sparse tree-like graphs. The RR graph is characterized by one connectivity parameter $c$, while the ER graph is characterized by the connection probability $p$. To keep the generated graph sparse enough in the ER case, we assume the probability is scaled as $p=\cbar/N$, yielding the mean degree $\cbar$. Furthermore, we also assume that the couplings of the teacher model 
have the same probability of taking both signs and the strength is constant: $|J^*_{i}|=\str>0$. The coupling strength $\str$ is assumed to be small enough to satisfy the paramagnet assumption of the teacher model. In particular, for the RR graph, the paramagnetic condition is 
\be 
(c-1)\tanh^2 \str < 1,
\ee 
while that of the ER one is 
\be
\cbar \tanh^2 \str < 1.
\ee
For readers interested in the derivation, please refer to~\cite{opper2001advanced,mezard2009information}. The cost function is fixed to that of PL in the following, as the simplest and commonly used case. The result of the RR graph case is shown below in \Rsec{RR graph}, and that of the ER graph is in \Rsec{ER graph}. For comparison, some numerical results on the square lattice are shown in \Rsec{Square lattice}, focusing on the approximation nature of the present theoretical results. Furthermore, as another common cost function, the so-called {\it interaction screening} (IS) method~\cite{vuffray2016interaction} belonging to the local learning class is examined and compared with PL.

Owing to the uniformity of the coupling strength, the strength of mean estimates $\{\bar{J}_i\}_{i\in \Omega}$ can also be set to a uniform value $|\bar{J_i}|=\bar{\str}=\Bias \str$, where the bias factor 
\be 
\Bias\equiv  \bar{\str}/\str,
\ee
is introduced. By the same reason, the marginal distribution can be simplified by again introducing the cavity field $h^*=\sum_{j\in \Omega}J_j^*s_j$ as
\be
\sum_{s_0,\V{s}_{\Omega}}
P(s_0,\V{s}_{\Omega}|\V{J}^*)(\cdots)
=\sum_{s_0}\int dh^* P_{\rm cav}(h^*|\V{J}^*)\frac{e^{s_0h^*}}{Z_0}(\cdots),
\Leq{reduction}
\ee 
where $Z_0=\int dh^*  P_{\rm cav}(h^*|\V{J}^*) 2\cosh h^*$. If the focused spin's connectivity is $c$, then the cavity field distribution becomes
\be
P_{\rm cav}(h^*|\V{J}^*)=
P_{\rm cav}(h^*|K,c)\equiv
\frac{1}{2^{c}}\sum_{k=0}^{c}
\binom{c}{k}
 \delta\lb h^*-\str(c-2k) \rb.
\Leq{P(h|c)}
\ee
Applying the reduction \NReq{reduction} in \Reqs{EOS_sparse}{Jbar} with replacement $\sum_{j\in \bar{\Omega}}\bar{J}_j s_j \to \Bias h^*$ in \Req{yhat} reduces the computation of mean estimates to that of the bias factor $\Bias$. The theoretically evaluated $\Bias$ was compared with that obtained by numerical experiments to check the validity of our theoretical treatment. 

The numerical computation of the order parameter $Q$ will be conducted below, but it has some delicate points. In our actual computations, the following procedure was adopted: From the generated teacher model we first compute the inverse correlation function $\lb C^{\bs 0}\rb^{-1}$ by the cavity formula \NReq{C_inverse} and numerically invert it to obtain $C^{\bs 0}$; then we introduce $\{ \hat{\Delta}_i=\hat{J}_{i}-\bar{J}_{i}\}_{i=1}^{N-1}$ from the learning result $\hat{\V{J}}$ and the mean estimate $\bar{\V{J}}$ which is obtained as $\bar{\V{J}}=\Bias\V{J}^*$ into which the theoretically evaluated value of $\Bias$ is inserted; finally we get a numerical value of $Q$ by $Q=\sum_{i,j}C^{\bs 0}_{ij}\hat{\Delta}_i\hat{\Delta}_j$. Although it is also possible to evaluate $C^{\bs 0}$ by the Monte Carlo (MC) sampling instead of using the formula \NReq{C_inverse}, this method is better for controlling fluctuations and reducing computational cost.

The outline of our numerical experiment is as follows. We first generate teacher model, next perform MC sampling, and finally choose a center spin and conduct learning by numerically minimizing the PL cost function~\NReq{estimator_PL}. The obtained estimator is used to compute relevant quantities such as RSS. As a result, there are some distinctive sources of fluctuation in the estimate, but we do not discriminate them below. The error bar is accordingly defined by the standard error coming from those fluctuations. The number of datasets used to compute the error bar is hereafter denoted as $N_{\rm set}$. More details of the numerical experiment are summarized in~\Rsec{Details of numerical}.

\subsection{RR graph case} \Lsec{RR graph}
In the case of the RR graph with connectivity $c$, using \Req{C_inverse} the trace of the inverse correlation function becomes 
\be
\frac{1}{N}\Tr{}C^{-1}=\frac{c}{1-\tanh^2\str}-c+1.
\ee
Substituting this in conjunction with the parameters obtained by \Reqss{EOS_sparse}{Jbar} into \Req{RSS_sparse2}, we obtain the RSS. Below, we compare these theoretical values with the numerically evaluated ones. 

We start by comparing the theoretical and numerical values of $\RSS$, $Q$, and $\Bias$. In \Rfig{RSS_OP-RR}, these quantities are plotted against $\alpha$ for $\str=0.2$ and $0.4$ at $N=200$ and $c=3$.
\begin{figure}[htbp]
\begin{center}
\includegraphics[width=0.32\columnwidth]{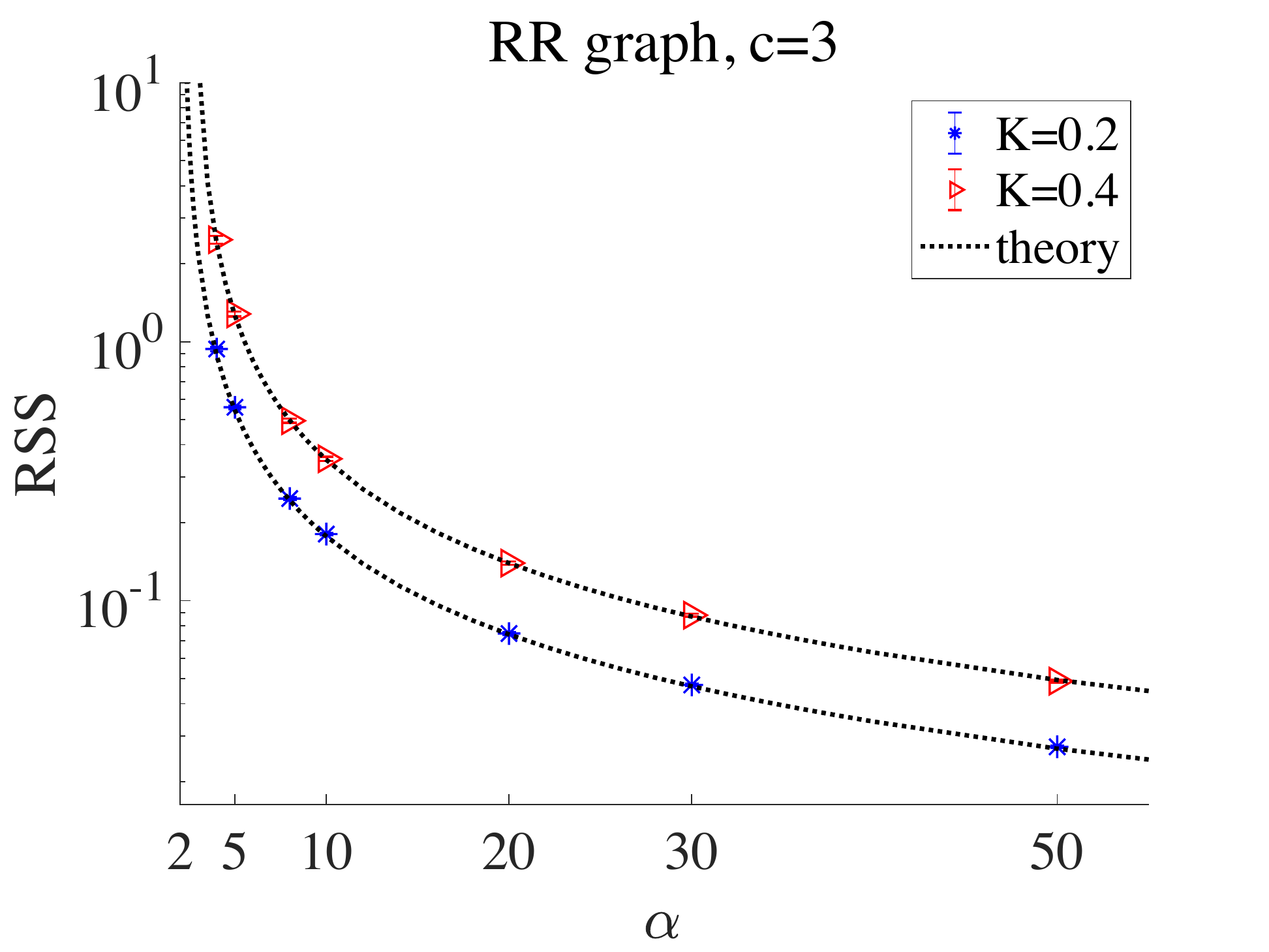}
\includegraphics[width=0.32\columnwidth]{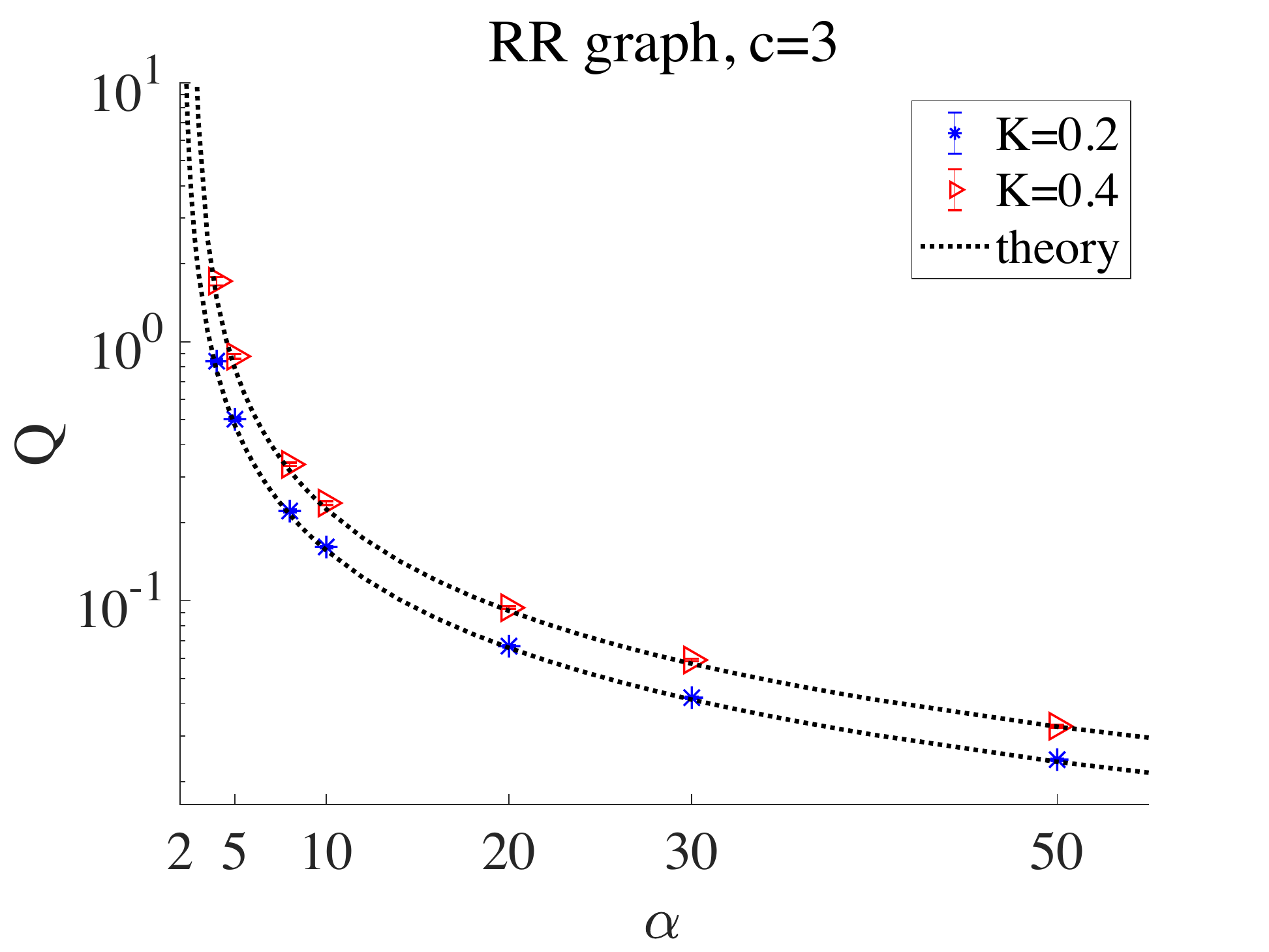}
\includegraphics[width=0.32\columnwidth]{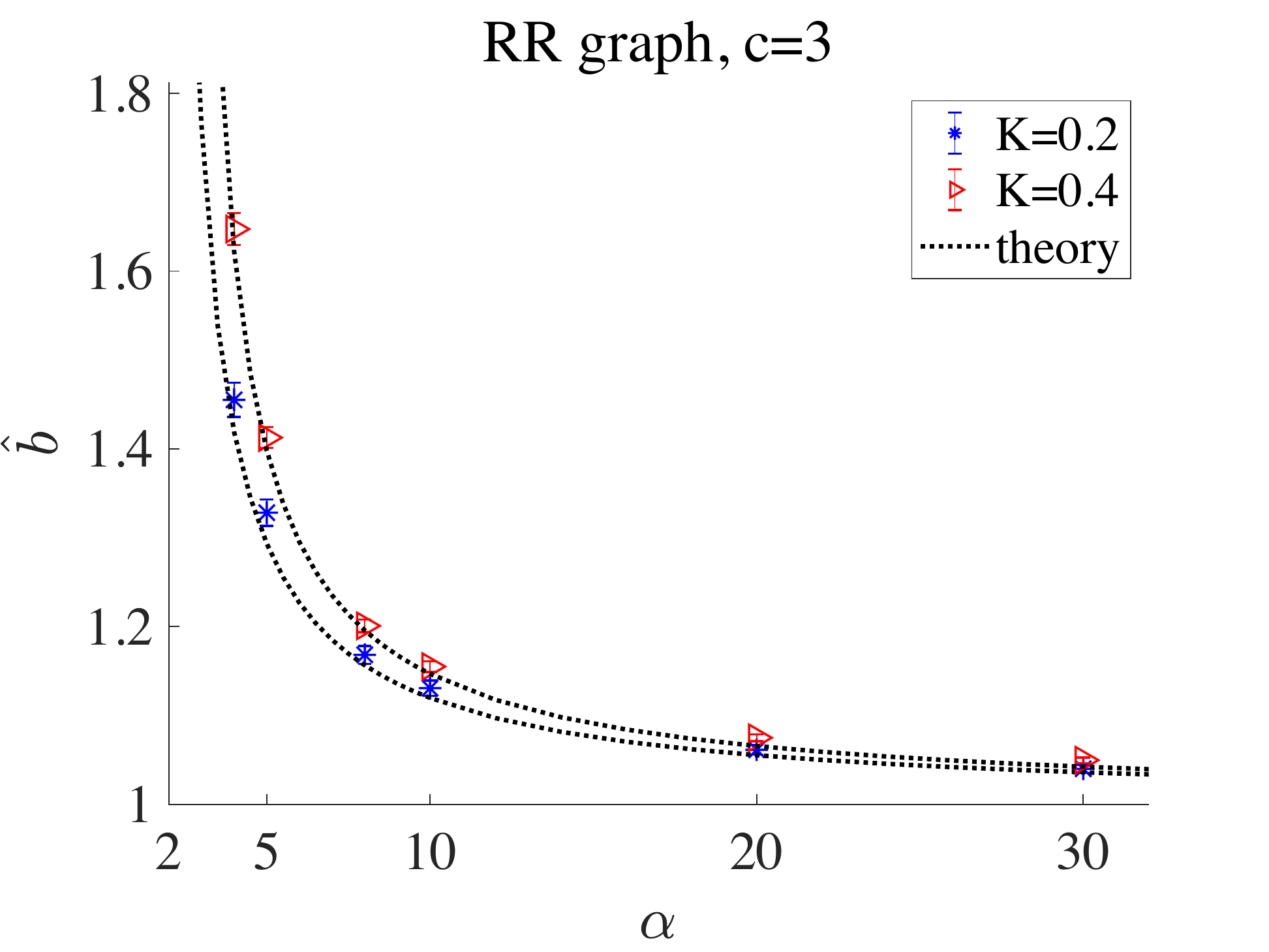}
\caption{Plots of $\RSS$ (left), $Q$ (middle), and $\Bias$ (right) against $\alpha$ for $\str=0.2$ and $0.4$ at $(N,c)=(200,3)$. Dotted lines and color markers are the theoretical and numerical values, respectively. The agreement between them is fairly good. The left and middle panels are plotted in the double log scale because $\RSS$ and $Q$ drastically diverge in the limit $\alpha \to 2$. The error bars obtained from $N_{\rm set}=100$ datasets are shown, although they tend to be comparable with the size of markers. 
}
\Lfig{RSS_OP-RR}
\end{center}
\end{figure}
In all the plots, the agreement between the theoretical (dotted lines) and numerical (color markers) results is fairly good, supporting the validity of our analytical treatment. 

Next, we consider the distributions of the estimators in \Rfig{Dist-RR}, which were normalized as probability distribution functions.
\begin{figure}[htbp]
\begin{center}
\includegraphics[width=0.32\columnwidth]{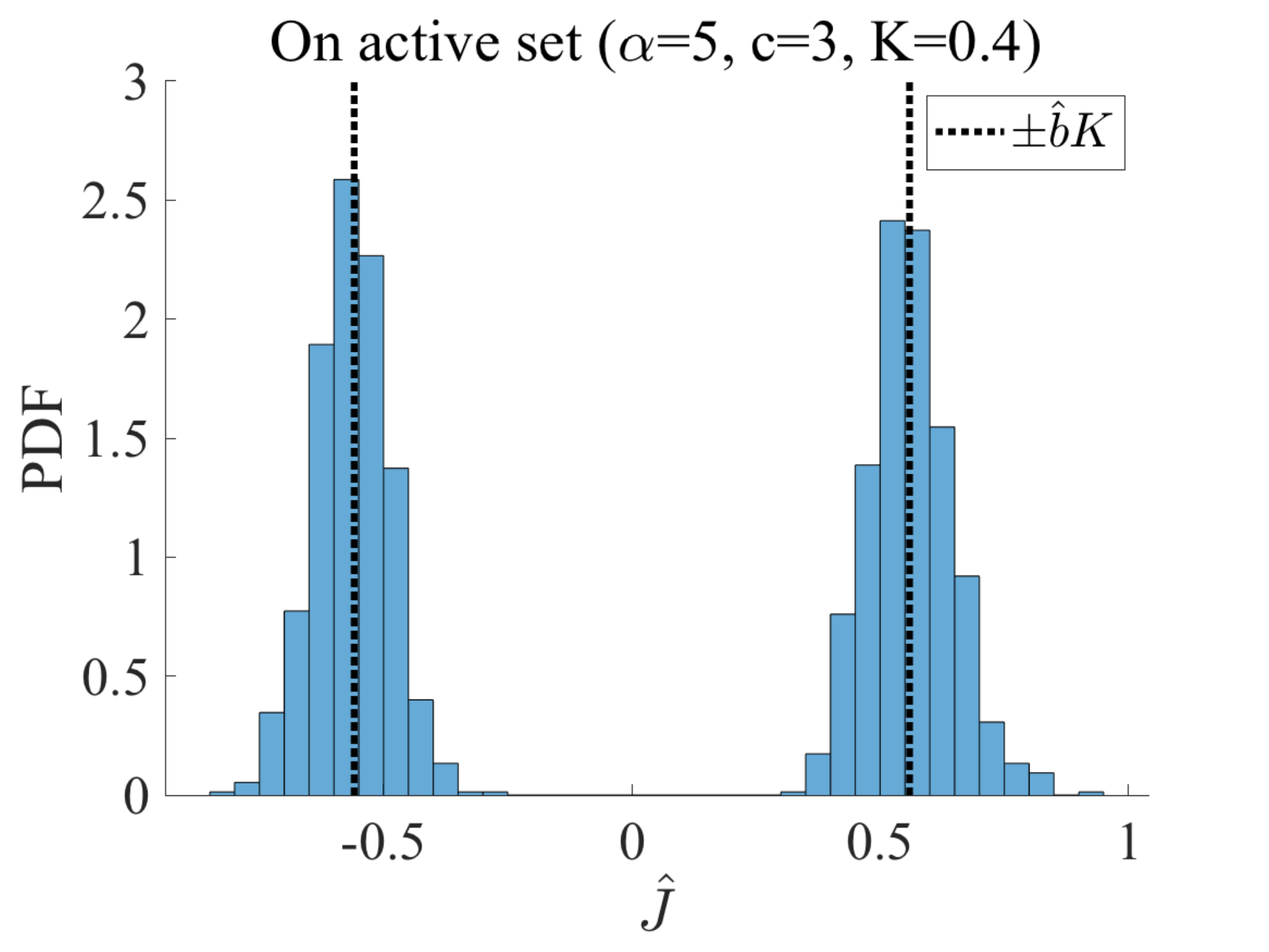}
\includegraphics[width=0.32\columnwidth]{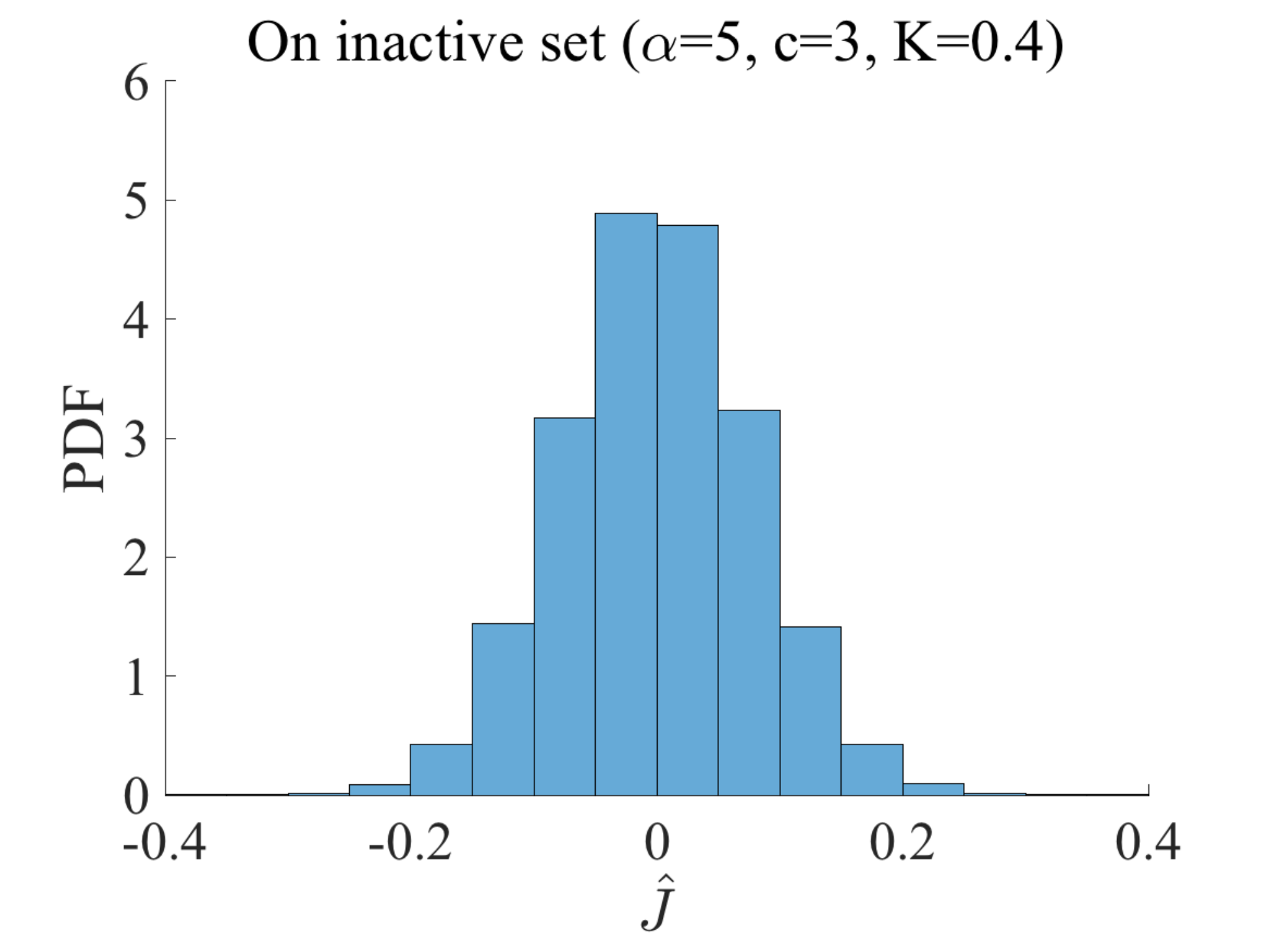}
\includegraphics[width=0.32\columnwidth]{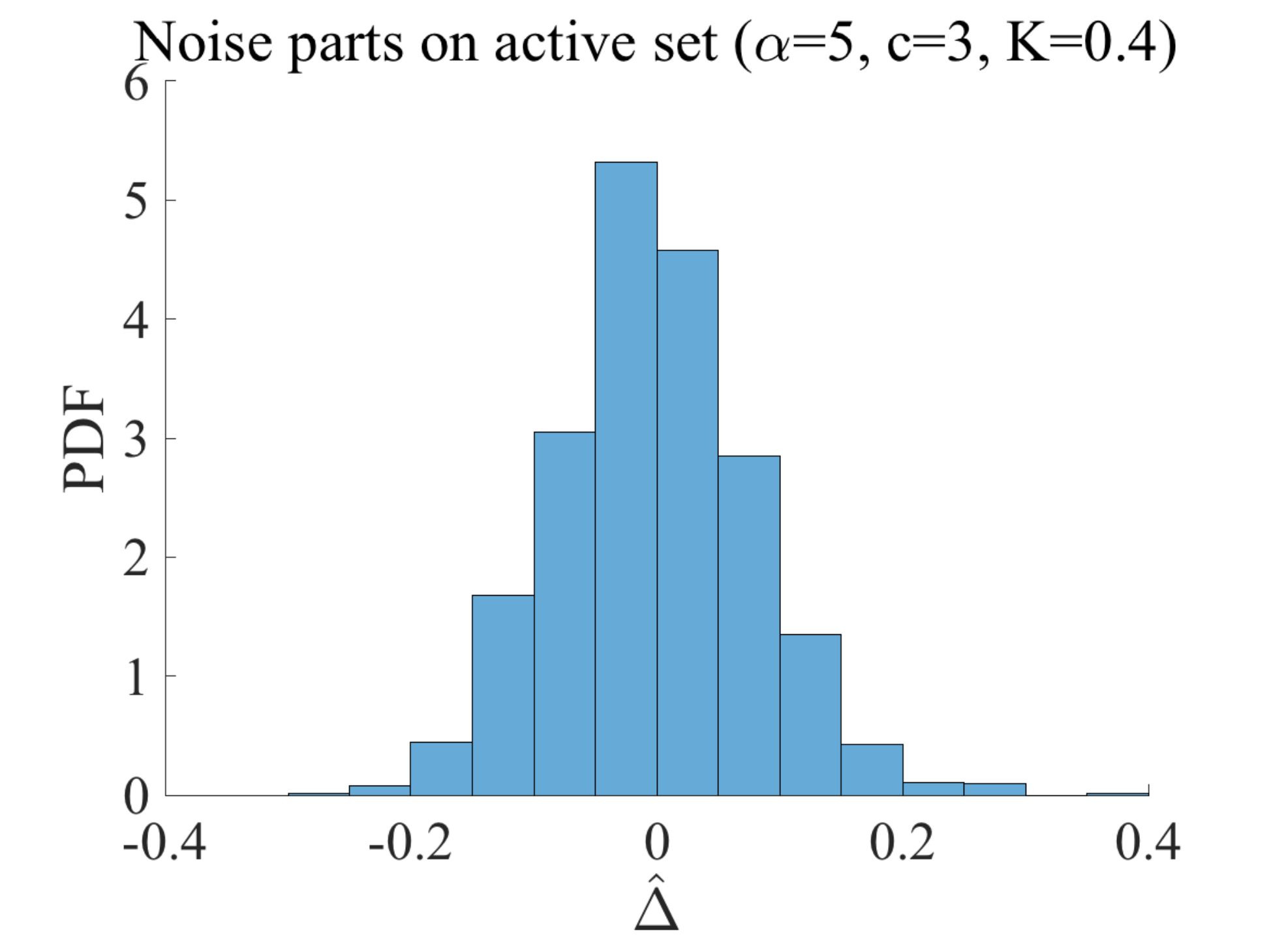}
\caption{Distribution of the estimators $\hat{\V{J}}$ on the active and inactive sets are given in the left and middle panels, respectively. The right panel is the distribution of the noise part on the active set, $ \{\hat{\Delta}_i=\hat{J}_{i}-\bar{J}_{i}\}_{i\in \Omega}$. The system parameters are $(N,\str,\alpha,c)=(200,0.4,5,3)$. The middle and right panels imply that the noise parts obey the zero-mean Gaussian distribution and have no discriminative difference between the active and inactive sets. Here, the histograms are generated from $N_{\rm set}=500$ datasets; from each dataset, the number of obtained estimators is $c=3$ for $\Omega$ while that for $\bar{\Omega}$ is $N-c-1=196$.  }
\Lfig{Dist-RR}
\end{center}
\end{figure}
The left panel is the distribution of the estimators on the active set $\Omega$. We can observe that two peaks are located around the theoretical prediction $\pm \Bias \str$. In the middle panel, the estimator distribution on the inactive set $\bar{\Omega}$ is shown, yielding a Gaussian-like distribution with zero mean. Similar behavior is observed for the noise part on the active set, $\{\hat{\Delta}_i=\hat{J}_{i}-\bar{J}_{i}\}_{i\in \Omega}$, the distribution of which is given in the right panel. Here, the mean estimates $\{\bar{J}_{i}\}_{i\in \Omega}$ are computed by multiplying the theoretically evaluated bias $\Bias$ by the true coupling $\{J_{i}^*\}_{i\in \Omega}$. These observations are again consistent with our theoretical analysis. 

Thirdly, we check the finite size effect. In \Rfig{FSE-RR}, against the system size $N$, the RSS and rescaled variance (multiplied by $N$) of the noise parts $\hat{\V{\Delta}}=\hat{\V{J}}-\bar{\V{J}}$ are plotted in the upper and lower panels, respectively.
\begin{figure}[htbp]
\begin{center}
\includegraphics[width=0.45\columnwidth]{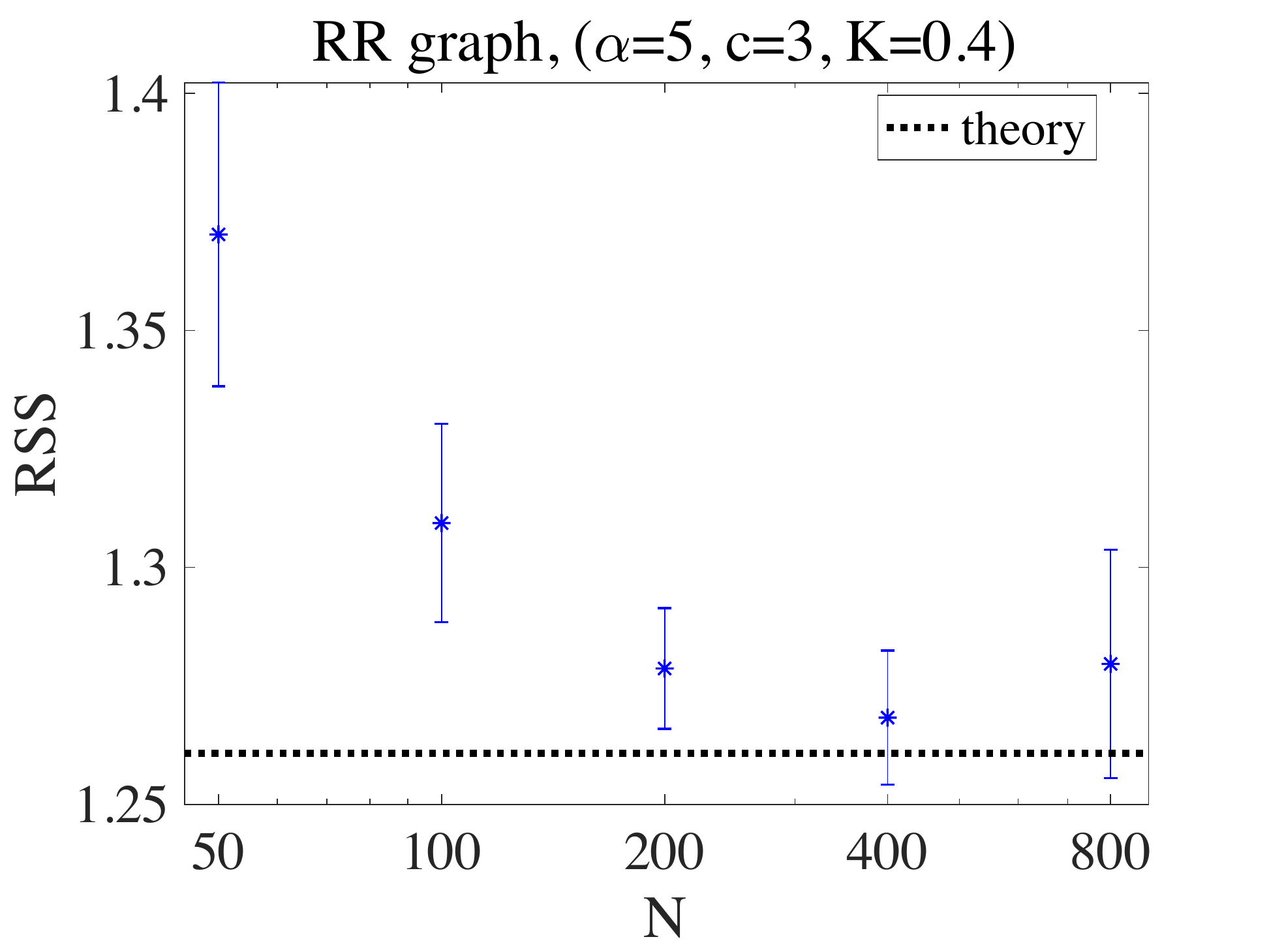}
\includegraphics[width=0.45\columnwidth]{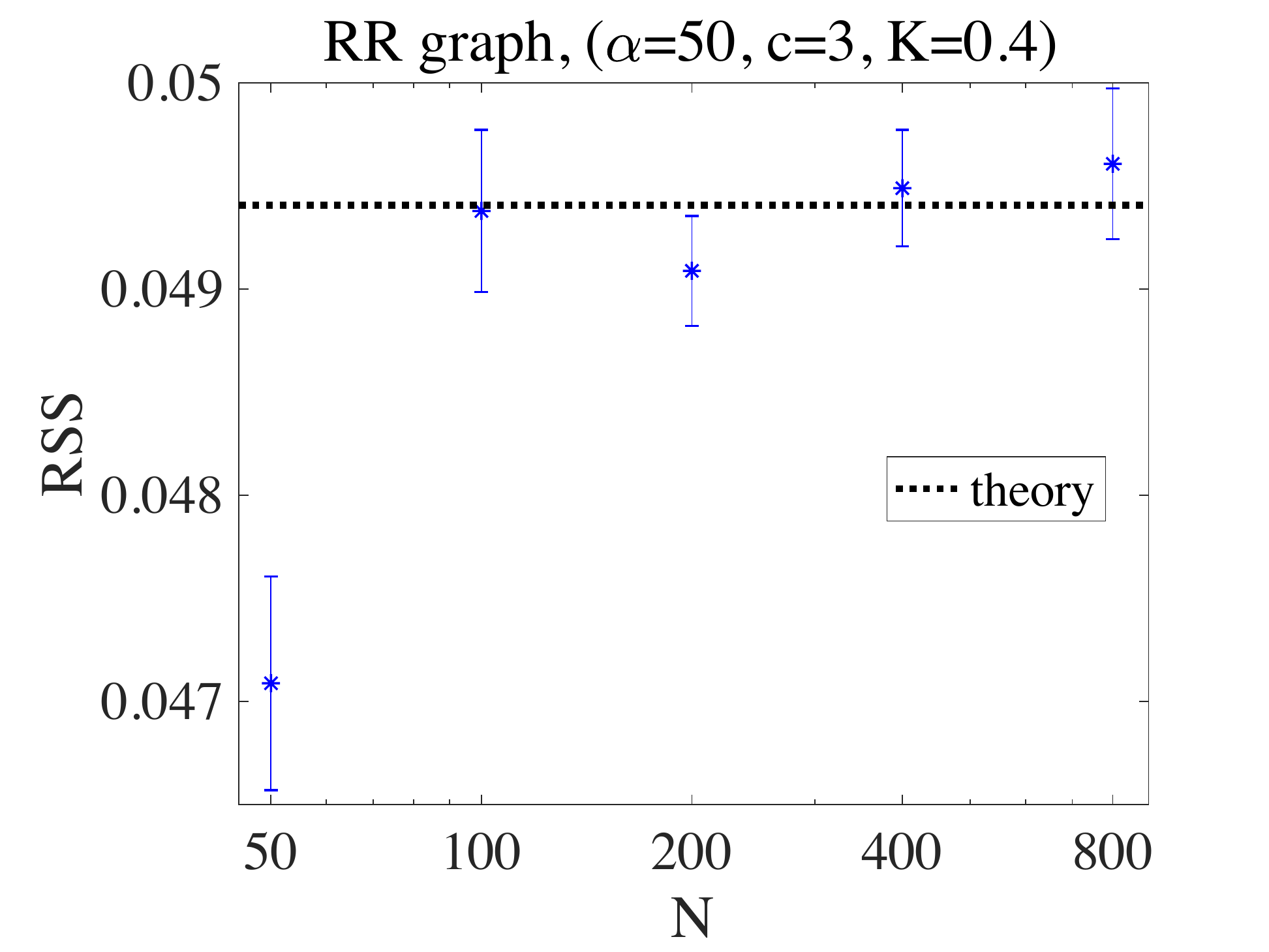}
\includegraphics[width=0.45\columnwidth]{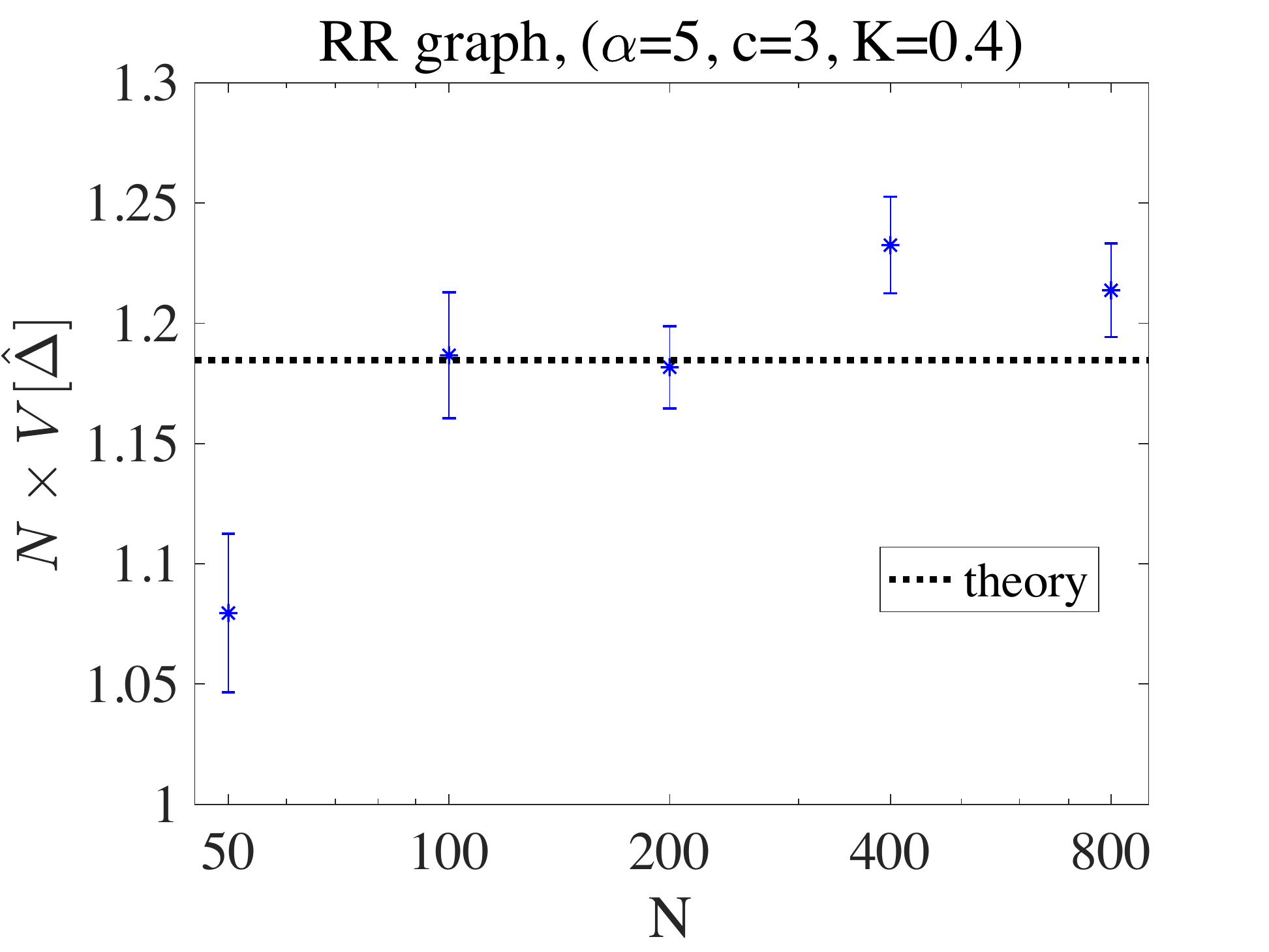}
\includegraphics[width=0.45\columnwidth]{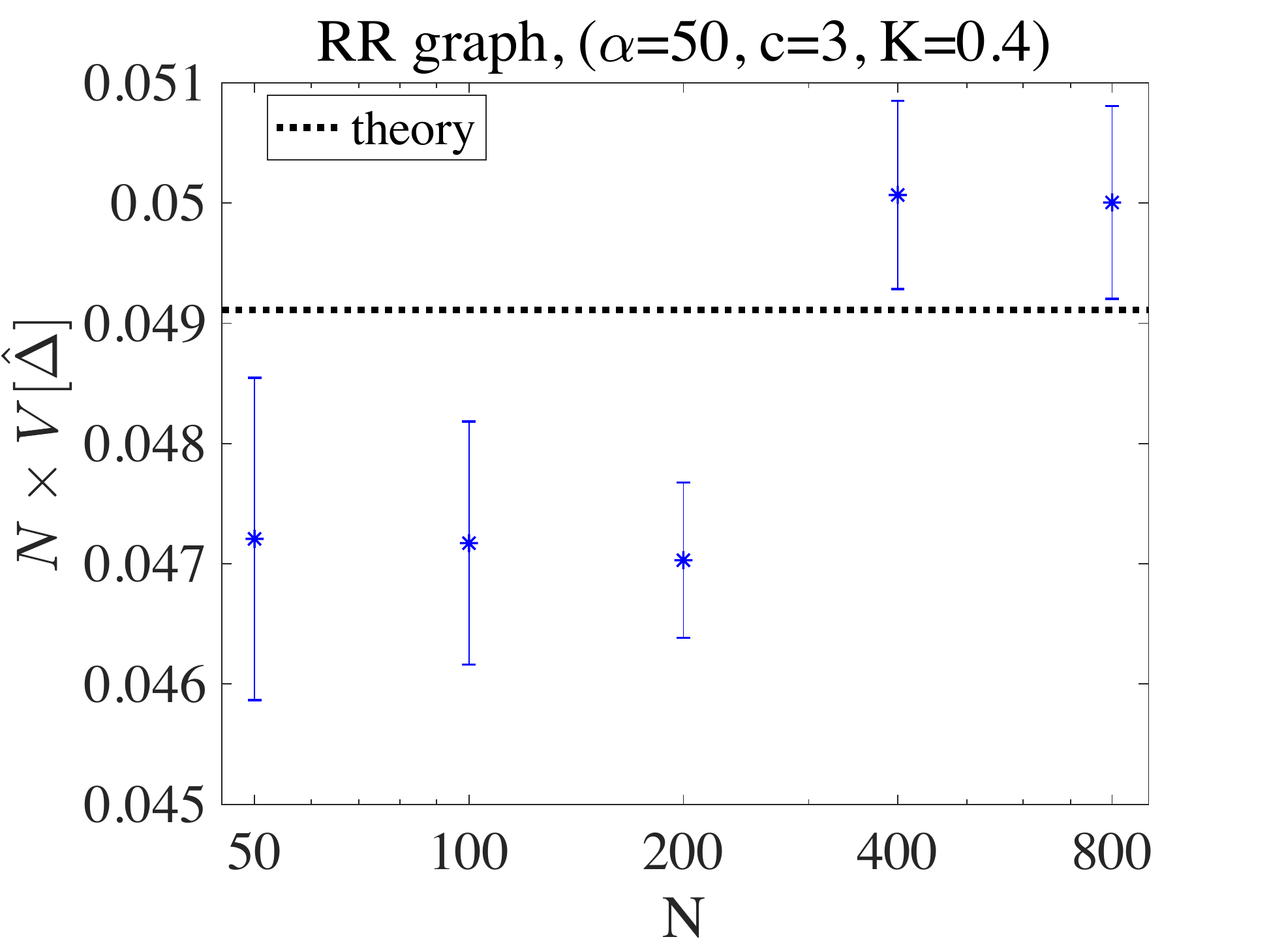}
\caption{(Upper): Plot of $\RSS$ against the system size $N$ for $\alpha=5$ (left) and $50$ (right) at $(\str,c)=(0.4,3)$. The black dotted lines denote the theoretical result and the markers are the numerical ones. The numerical results tend to converge with the theoretical results as the system size grows, although the finite size effect seems to be different between the left and right panels. The error bar is obtained from $N_{\rm set}=500$ datasets for $N=50$--$200$, $N_{\rm set}=400$ for $N=400$, and $N_{\rm set}=50$ for $N=800$.
(Lower): The rescaled variance (multiplied by $N$) of the noise part $\hat{\V{\Delta}}=\hat{\V{J}}-\bar{\V{J}}$ is plotted against the system size $N$. The parameters are the same as those of their counterparts in the upper panels. Although in this closeup scale there is a small gap between the numerical and theoretical results within the one standard error, this gap  can be eliminated by taking a larger number of samples. Here, the error bar was obtained using the bootstrap method by considering each realization and component of $\hat{\V{\Delta}}$ as i.i.d.. }
\Lfig{FSE-RR}
\end{center}
\end{figure}
Although the finite size effect behaves in different ways depending on the parameters and quantities, we can see that the numerical results (markers) fairly matched the theoretical values (black dotted lines) as the system size is large. Here, the rescaled variance corresponds to the quantity $Q\Tr{}\lb C^{\bs 0}\rb^{-1}/N$ in our theoretical computation, which is consistent with \Req{RSS_sparse2}. These results again confirm the validity of our computations. 

Finally, we have some noteworthy remarks. The results shown in \Rfigs{Dist-RR}{FSE-RR} imply the possibility of an efficient method of debiasing and hypothesis testing. The bias factor $\Bias$ can be computed from our analytical result, and hence we can debias our estimator in an efficient manner. The residual after debiasing $\hat{\V{\Delta}}$ is considered to obey a Gaussian distribution, as shown in \Rfig{Dist-RR}, and is supported by our analytical computations in \Rsec{Computations}. Thus, we can efficiently compute the P-value according to the standard hypothesis testing method, enabling us to judge the relevance of the estimated couplings. Moreover, in the thermodynamic limit $N\to \infty$, we can show that the perfect reconstruction of the teacher's network is possible for any $\alpha >2$. To do so, we need to evaluate the probability of getting false positives in the estimator. To control false positives, we introduce a constant threshold value $\str_{\rm th}(>0)$, and consider estimated couplings with absolute values less than $\str_{\rm th}$ as negligible and set to zero; we independently repeat this procedure for all $i=1,\cdots,N$. Let us evaluate the probability of successfully screening out false positives using this method. The observations so far imply, on the inactive set $\bar{\Omega}$, that the estimator behaves as
\be
\hat{J}_{i}\sim \mathcal{N}\lb0,\frac{\sigma_i^2}{N}\rb,~(\forall i\in \bar{\Omega}),
\ee
where $\sigma_i^2(=O(1))$ is the rescaled variance of the estimate, with relation $(1/N)\sum_{i\in \bar{\Omega}}\sigma_i^2 \approx Q\Tr{}\lb C^{\bs 0}\rb^{-1}/N$. Hence, the probability of successfully screening out these estimators on $\bar{\Omega}$ is 
\be
&&
\prod_{i\in \bar{\Omega}}{\rm Prob}\lb |\hat{J}_{i}|< \str_{\rm th} \rb
=
\prod_{i\in \bar{\Omega}}\lb 1-2\int_{\sqrt{\frac{N}{\sigma_i^2}}\str_{\rm th} }^{\infty}dz~\frac{e^{-\frac{1}{2}z^2}}{\sqrt{2\pi}}\rb
\no \\ &&
\approx 
\prod_{i\in \bar{\Omega}}\lb 1
-\frac{2}{\sqrt{2\pi}}
\frac{   e^{-\frac{1}{2}  \frac{N}{\sigma_i^2}\str_{\rm th}^2 } }{  \sqrt{\frac{N}{\sigma_i^2}}\str_{\rm th} }\rb
\to 1,~(N\to \infty).
\ee
The second approximate equality comes from the asymptotic formula of the integral, which can be justified for large $N$. The last limiting behavior holds as long as $\sigma_i$ is bounded from above, because the exponential factor $e^{-\frac{1}{2} \frac{N}{\sigma_i^2}\str_{\rm th}^2 }$ decays fast enough compared with the number of products $|\bar{\Omega}|=N-c-1$. Hence, we can completely suppress the false positives in the limit $N\to \infty$. Meanwhile, we also desire to accurately reproduce the presence of couplings on $\Omega$. This can be done by tuning the threshold value $\str_{\rm th}$ as smaller than the true coupling strength $\str$ (the mean estimates $\bar{\V{J}}$ are larger than $\str$ in the absolute value). In practical situations, we do not know the true coupling strength $\str$ in advance, and thus it is nontrivial to correctly tune $\str_{\rm th}$. In such cases, it may be better to tune $\str_{\rm th}$ by monitoring the distribution of estimators such as \Rfig{Dist-RR}, and to find a value that effectively separates the modes of distribution. The present theoretical analysis supports this process by manifesting the behavior of estimators in the limit $N\to \infty$.

\subsection{ER graph case} \Lsec{ER graph}
For the ER graph with connection probability $p=\cbar/N$, the evaluation of the order parameters and related quantities is slightly more complex than the RR case because of the distributed nature of the connectivity. In the thermodynamic limit, the distribution of connectivity $c$ in the ER graph obeys the Poisson distribution:
\be
P_{\rm po}(c|\cbar)=e^{-\cbar}\frac{\cbar^{c}}{c!}.
\ee
The trace of the inverse correlation function fortunately becomes simple in the limit:
\be
\frac{1}{N}\Tr{}C^{-1}
\xrightarrow[]{N\to \infty}
\sum_{c=0}^{\infty}
\lb 
\frac{c}{1-\tanh^2\str}-c+1
\rb
P_{\rm po}(c|\cbar)
=
\lb 
\frac{\cbar}{1-\tanh^2\str}-\cbar+1
\rb.
\ee
When focusing on spin $i$ with connectivity $c_i$ in the ER graph, its associated order parameters are computed by \Req{EOS_sparse} with $P(h^*|c_i)$ defined in \Req{P(h|c)}, and the RSS is given by 
\be 
\RSS_i(c_i)=
(1-\Bias(c_i))^2\str^2
+
Q(c_i)\lb \frac{\cbar}{1-\tanh^2\str}-\cbar+1\rb.
\ee
This explicit dependence of the order parameter on $c_i$ is the complex  point of the ER case. Upon realizing this property, we can easily compute the mean RSS for the whole network, which is written as 
\be
\RSS_{\rm mean}=\frac{1}{N}\sum_{i=1}^{N}\RSS_i
\xrightarrow[]{N\to \infty}
\sum_{c=0}^{\infty}
\lbb 
(1-\Bias(c))^2\str^2
+
\lb \frac{\cbar}{1-\tanh^2\str}-\cbar+1\rb
Q(c)
\rbb
P_{\rm po}(c|\cbar).
\Leq{RSS_whole_ER}
\ee
These provide explicit formulas of the RSSs for the ER case.

As an interesting departure from the RR case, we here examine the connectivity dependence of our quantities of interest. The plots of $\RSS(c), Q(c)$, and $\Bias(c)$ at $(N,\alpha,\cbar,\str)=(400,10,4,0.4)$ are given in \Rfig{RSS_OP-ER}.
\begin{figure}[htbp]
\begin{center}
\includegraphics[width=0.32\columnwidth]{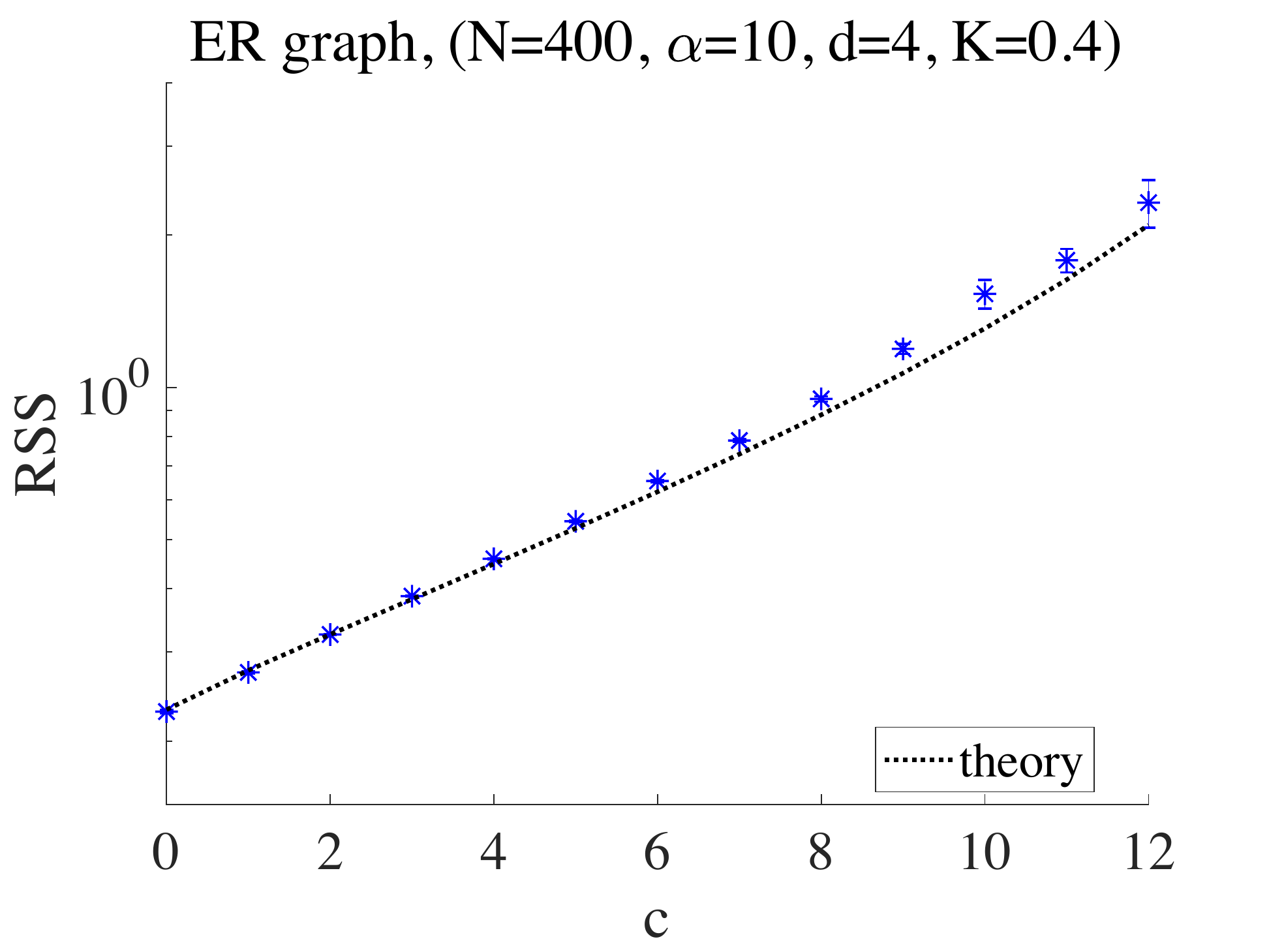}
\includegraphics[width=0.32\columnwidth]{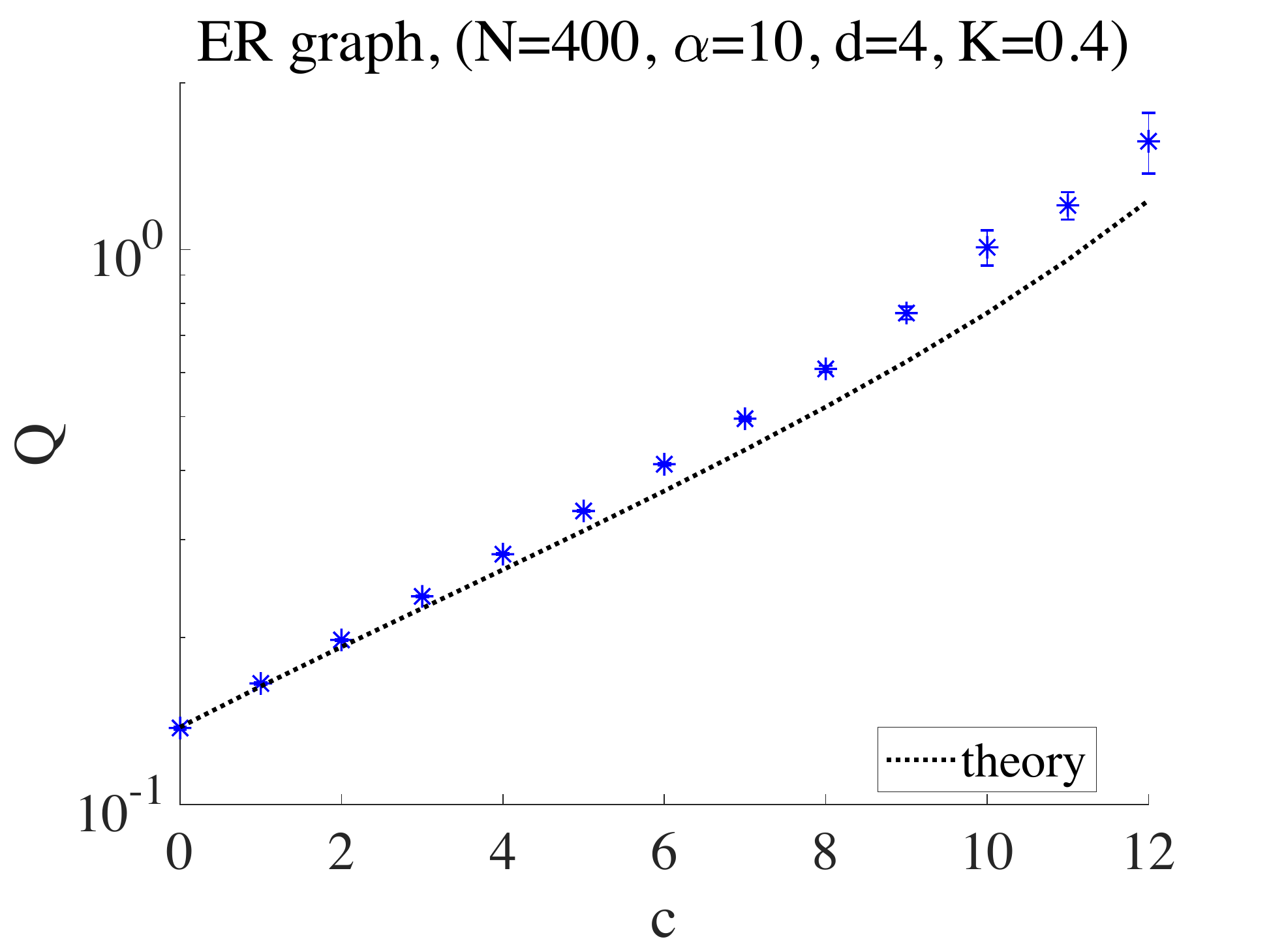}
\includegraphics[width=0.32\columnwidth]{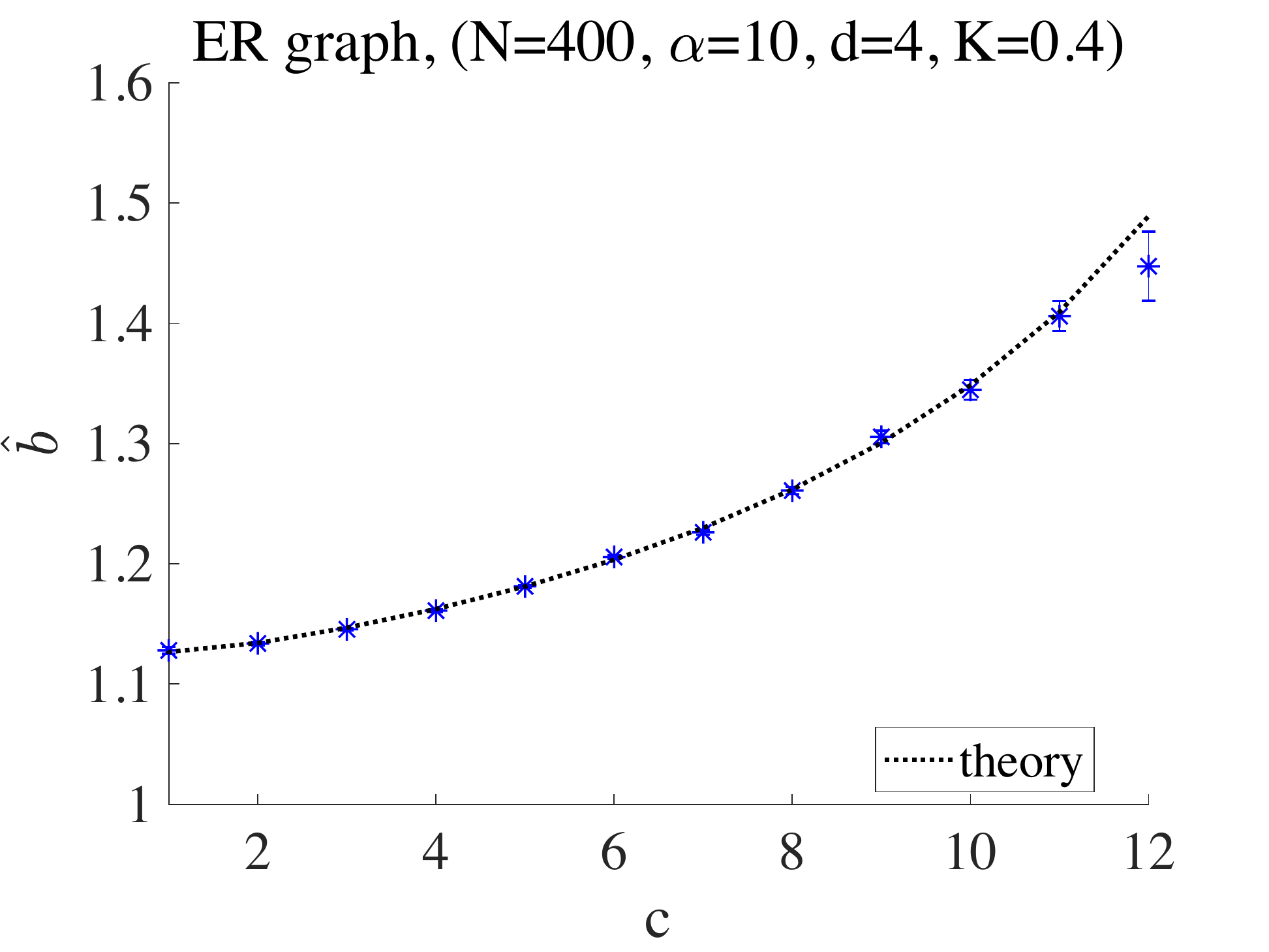}
\caption{Plots of $\RSS$ (left), $Q$ (middle), and $\Bias$ (right) against $c$ at $(N,\alpha,\cbar)=(400,10,4)$. Black dotted lines and markers are the theoretical and numerical values, respectively; the different colors correspond to different $\str$. The agreement between them is fairly good. 
}
\Lfig{RSS_OP-ER}
\end{center}
\end{figure}
In this experiment, we generated ten different ER networks, performed two independent MC samplings, and conducted learning for all $i=1,\cdots,N$. The error bars were placed using the obtained datasets, and thus $N_{\rm set}$ varied among different connectivity $c$. The agreement between the theoretical and numerical results is fairly good, supporting our theoretical result. Although a slight deviation at large $c$ in $\RSS(c)$ and $Q(c)$ was observed, this was presumably attributed to the finite size effect, which increased at large $c$ because of insufficient system size for generating nodes with large $c$. We have tried to control this deviation but found it is difficult to conduct experiments of sufficiently large systems in reasonable time: The generation probability of node with, say, $c=13$ can be estimated as $P_{\rm po}(13|4)\approx 2\times 10^{-4}$, and hence for stably generating networks with such large degree nodes we need at least $N\approx 5000$, which is too much in our experiment. We thus leave this problem untouched. 

We also computed the mean RSS \NReq{RSS_whole_ER} for the whole network. The theoretical value is $\RSS_{\rm mean}=0.4780$, while the present experimental value is $\RSS_{\rm mean}=0.4907\pm 0.0041$. The slight difference between these is again attributed to the finite size effect. Here, the theoretical value was obtained by taking the sum of \Req{RSS_whole_ER} up to $c=20$; the effect of this truncation was found to be small.

\subsection{Square lattice case for comparison} \Lsec{Square lattice}
The cavity method in direct problems is known to yield good approximations even for loopy graphs, when correlations among spins are weak; it is sometimes referred to as Bethe approximation. Here, we examined this approximation nature of the present theoretical computation of inverse problems. To this end, we compared our theoretical result for $c=4$ with the simulation result on the square lattice with periodic boundary condition. To avoid possible complexity due to frustration, the present teacher couplings were assumed to be all positive and constant, $J_{i}^*=\str >0, (i\in\Omega)$. 

In \Rfig{RSS_OP-RR}, we plotted $\RSS$ and $\Bias$ against $\alpha$ for $\str=0.2$ on the square lattice of size $20\times 20$, in comparison with our theoretical result (dotted line) computed with the assumption of the tree-like network structure. 
\begin{figure}[htbp]
\begin{center}
\includegraphics[width=0.45\columnwidth]{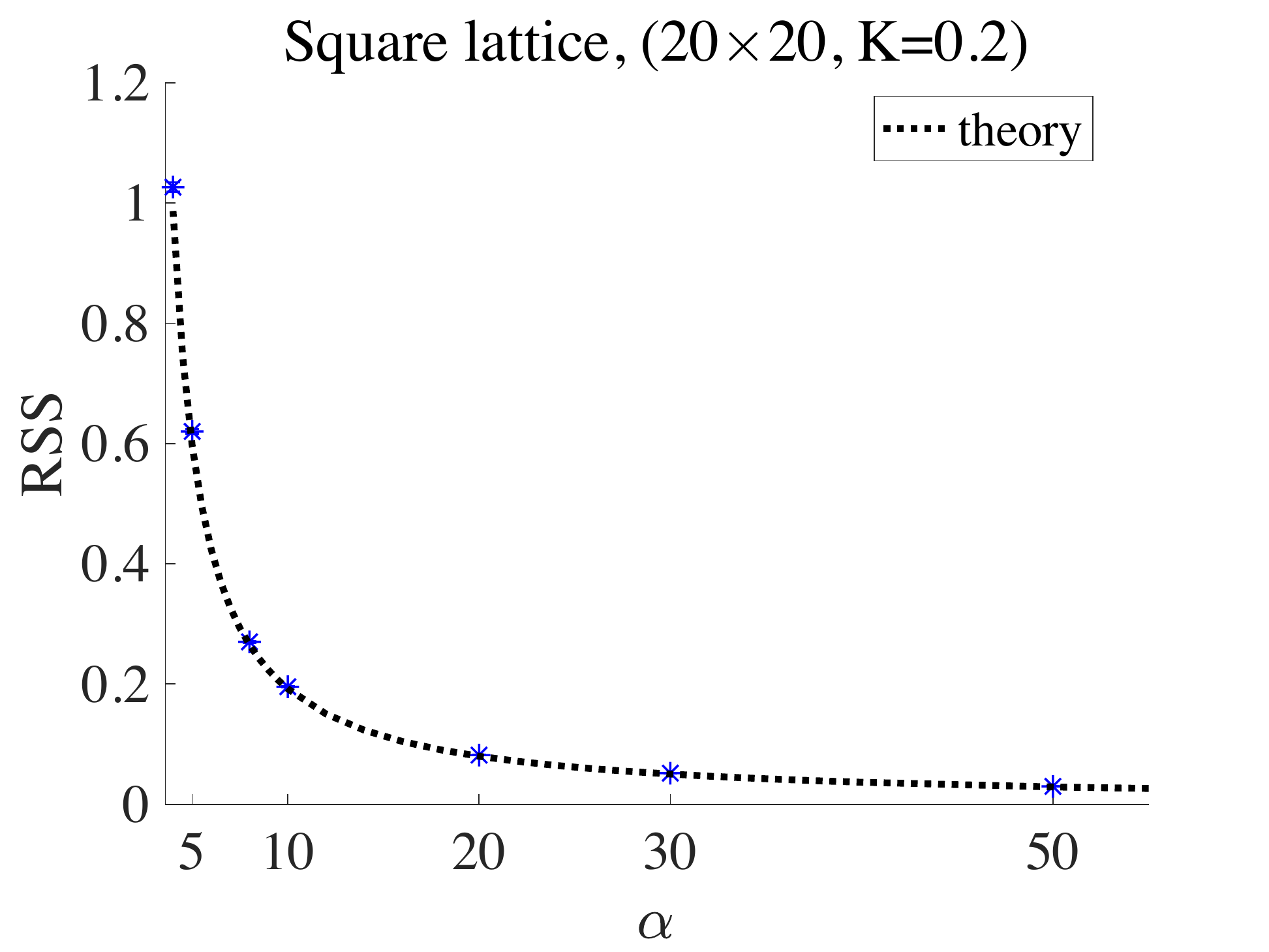}
\includegraphics[width=0.45\columnwidth]{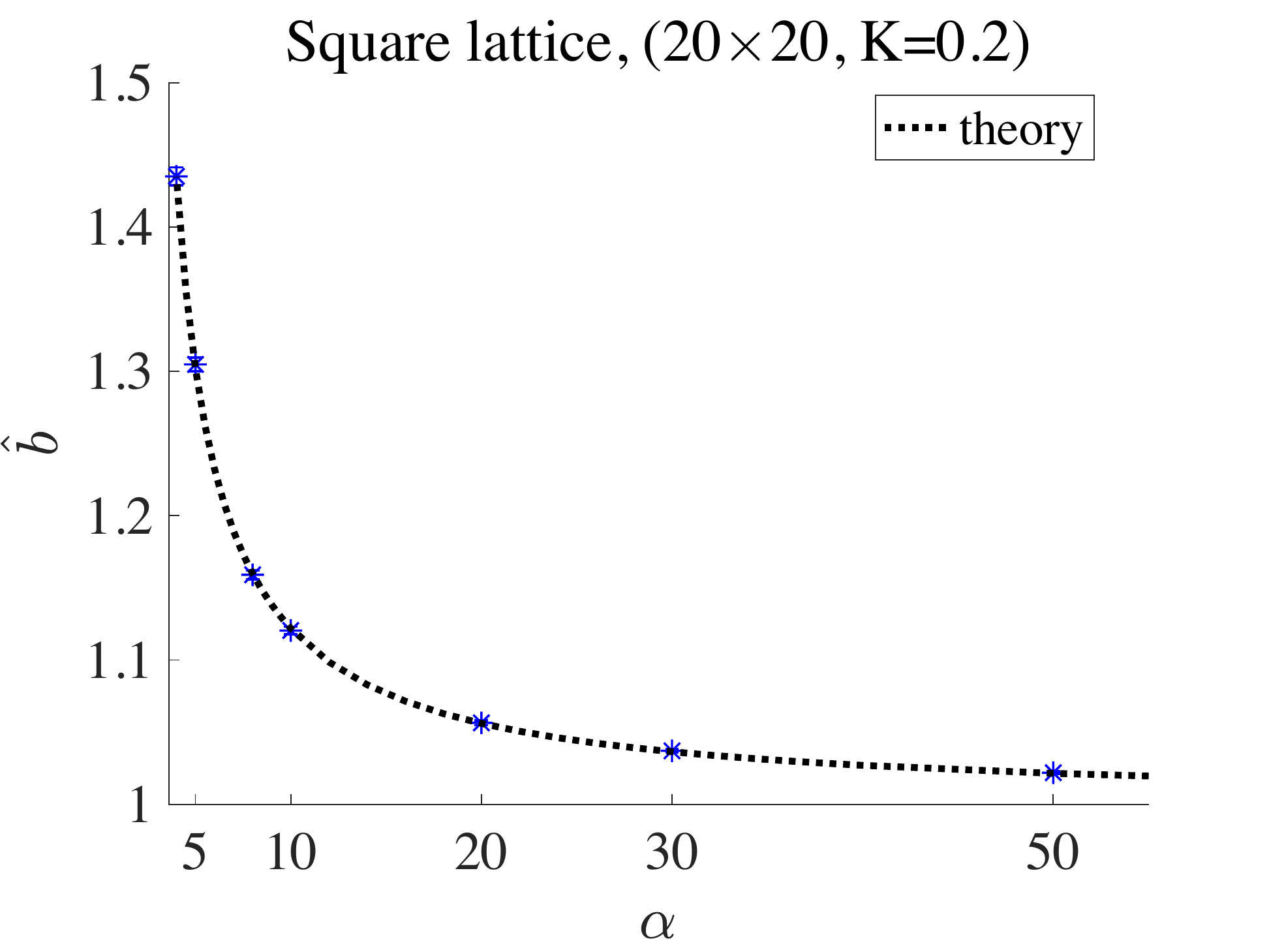}
\caption{Plots of $\RSS$ (left) and $\Bias$ (right) against $\alpha$ for $\str=0.2$ on the square lattice of size $20\times 20$. For comparison, the theoretical results derived by assuming the tree-like structure of the coupling network are plotted as the dotted lines. The agreement between the markers (numerical results) and lines is fairly good. The error bars obtained from $N_{\rm set}=400$ datasets are shown.
}
\Lfig{RSS_OP-RR}
\end{center}
\end{figure}
The agreement between the theoretical and numerical results is fairly good. This indicates that our theoretical result can be a good approximation even for loopy graphs. 

Another interesting phenomenon for loopy graphs is the possible presence of bias in the estimated couplings for spins in $\bar{\Omega}$, as discussed in \Rsec{Applicable range}. To examine this, in the upper panels of \Rfig{bias-square} we show the distributions of the coupling estimates corresponding to the next nearest neighbors (NNN) from the center spin $s_0$ in the teacher model for the square lattice (left) and for the RR graph with $c=4$ (right).  
\begin{figure}[htbp]
\begin{center}
\includegraphics[width=0.45\columnwidth]{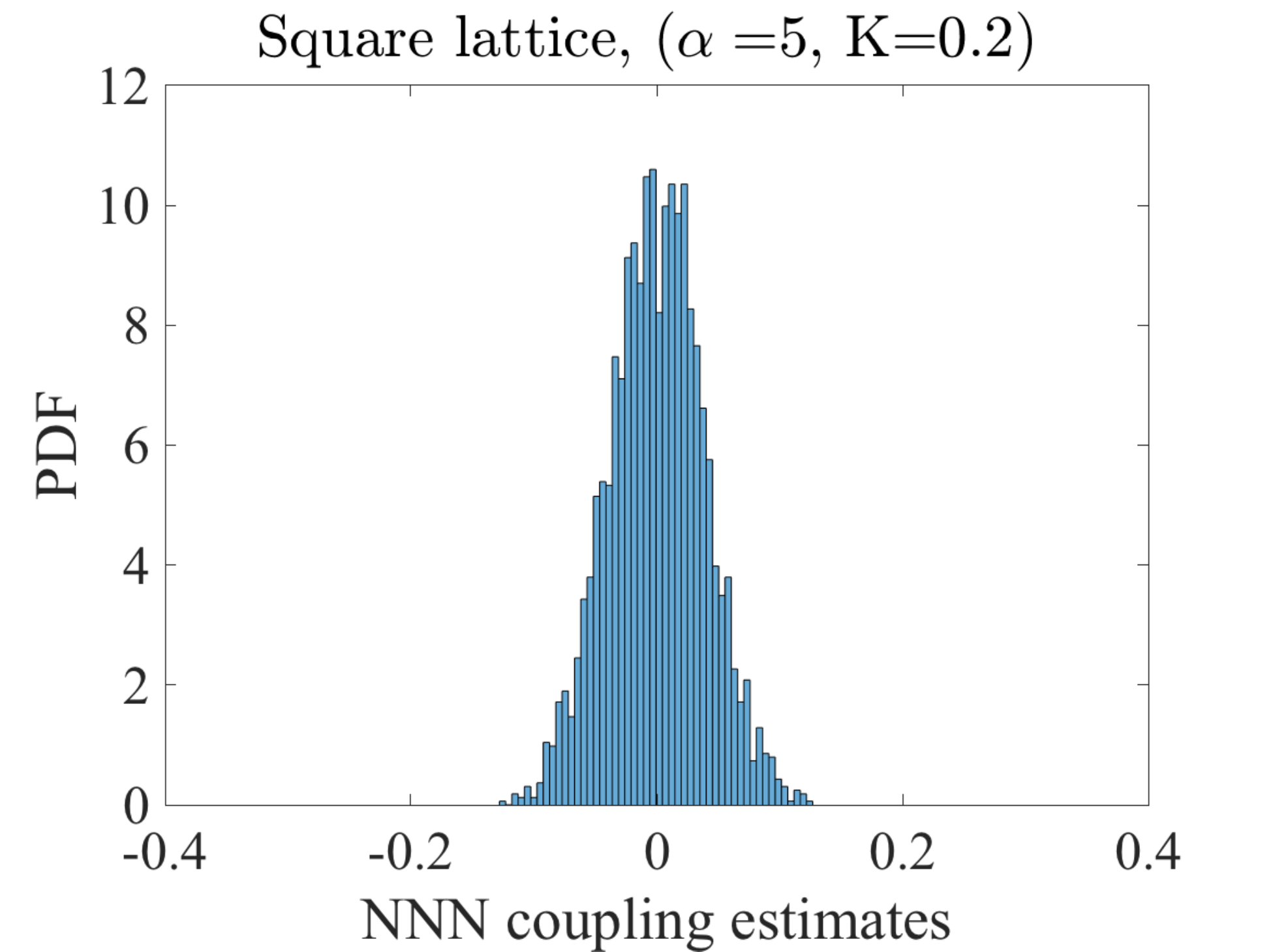}
\includegraphics[width=0.45\columnwidth]{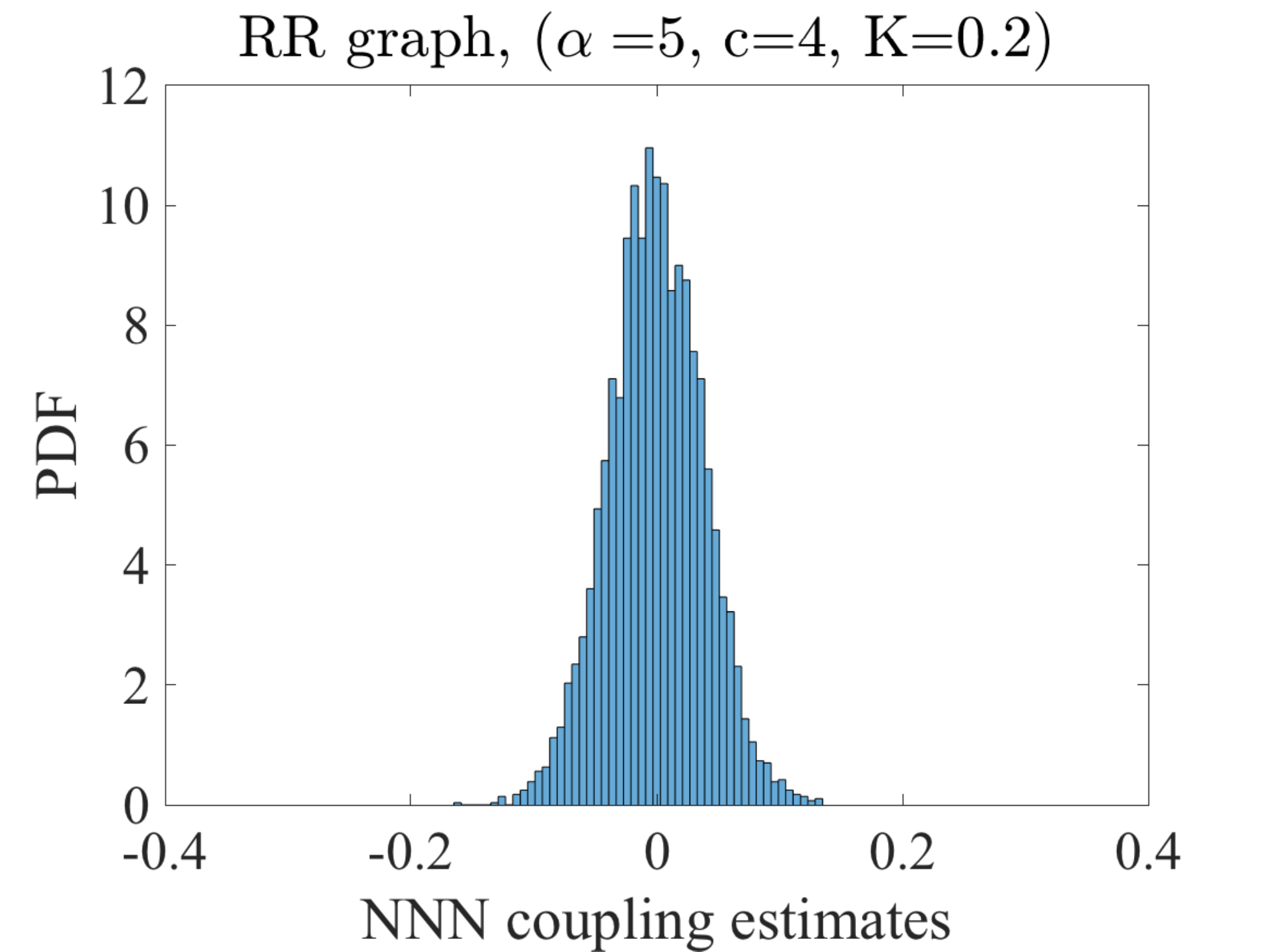}
\includegraphics[width=0.45\columnwidth]{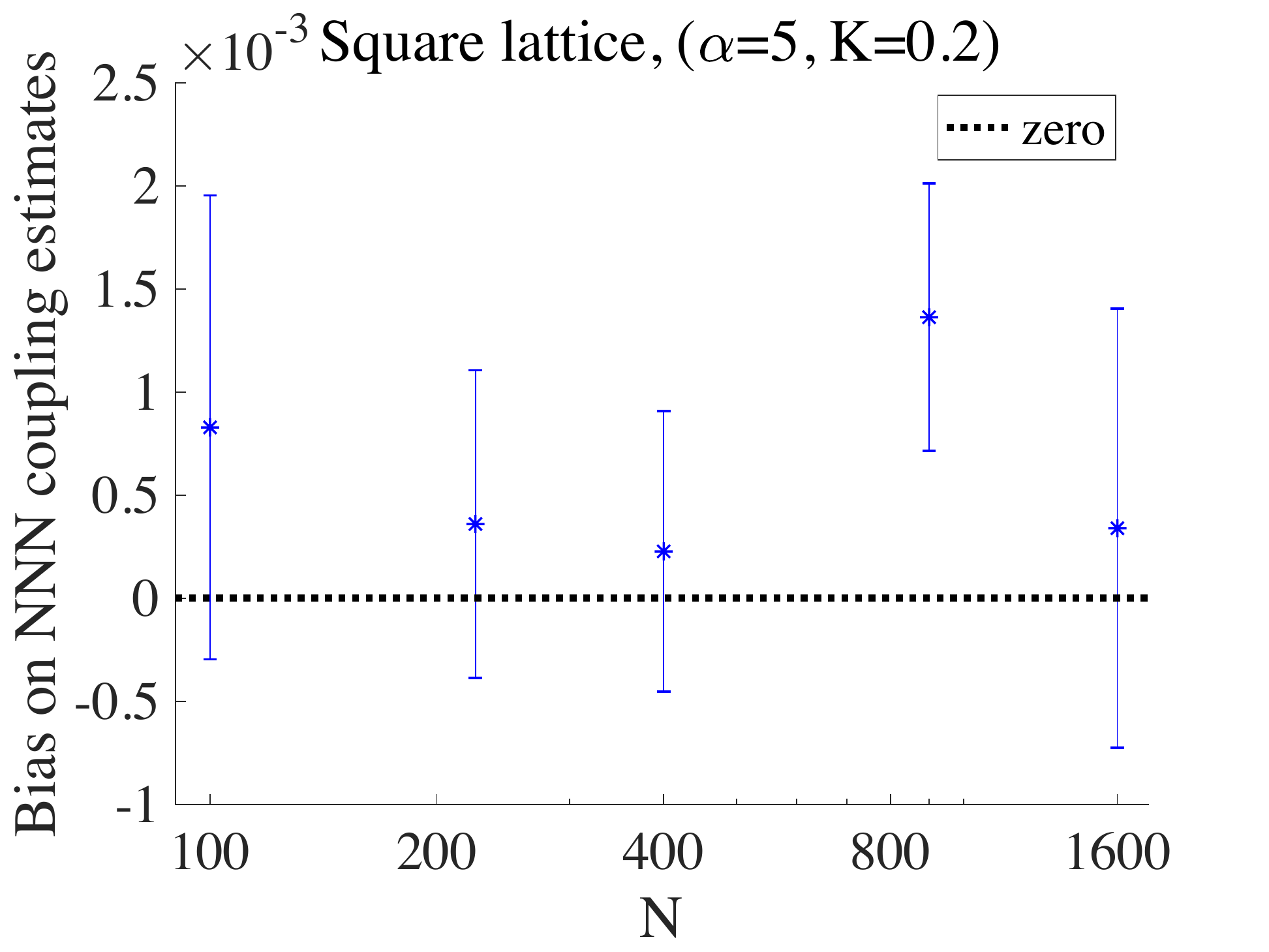}
\includegraphics[width=0.45\columnwidth]{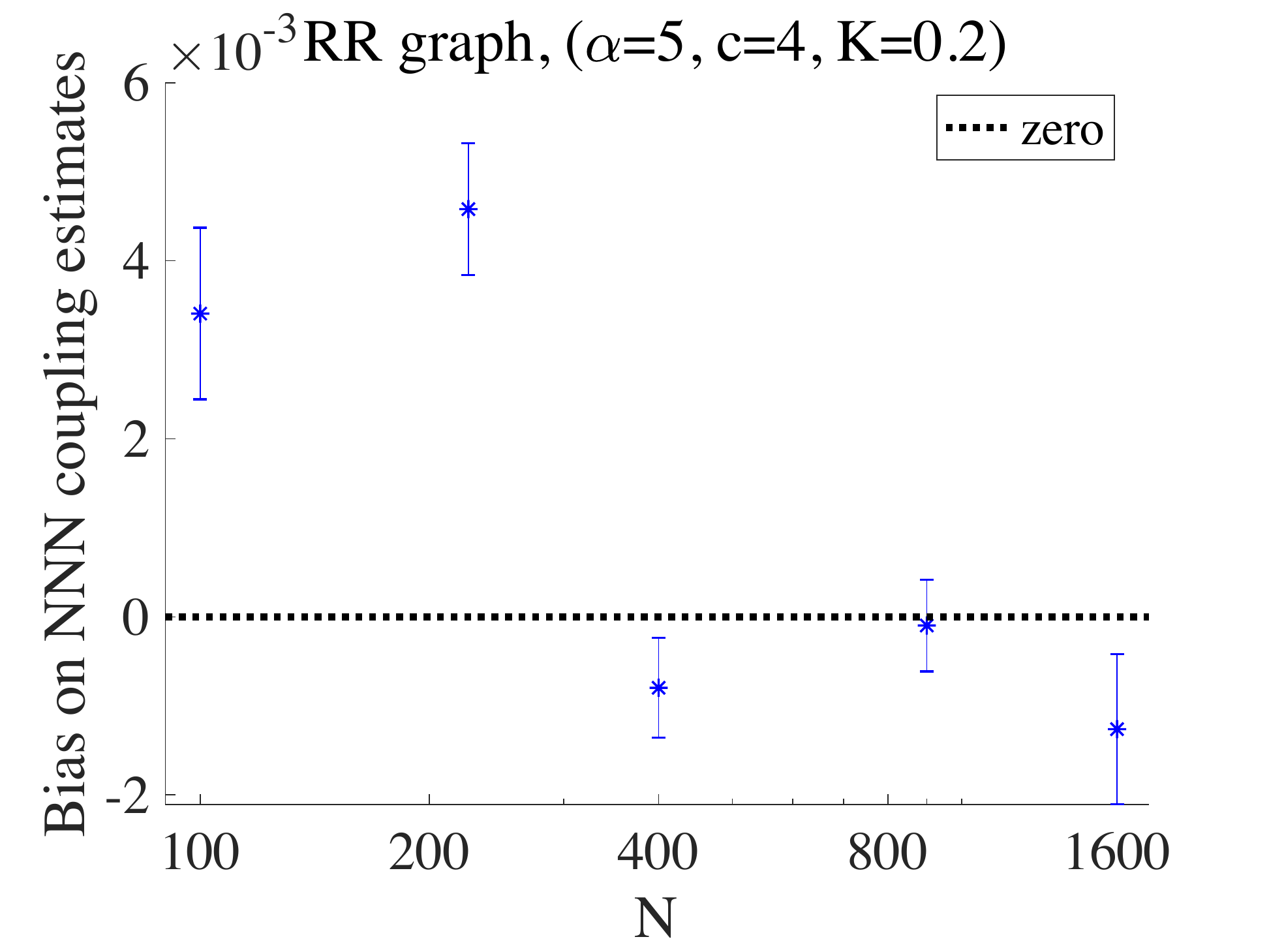}
\caption{(Upper): Distributions of the NNN estimators for the $20\times 20$ square lattice (left) and for the RR graph with $(N,c)=(400,4)$. In both  cases, other parameters are set to be $(\str,\alpha)=(0.2,5)$ and $N_{\rm set}=400$. No clear positive/negative tendency is observed in both cases. (Lower). Plots of the mean of the NNN estimate distribution against the system size for the square lattice (left) and RR graph (right). The other parameters are similar to those of the corresponding upper panels. The means are quite small, and no clear deviation from zero is observed. The dataset sizes are $N_{\rm set}=600,600,400,200,40$ for $N=100,225,400,900,1600$, respectively. The error bars are obtained using the bootstrap method. 
} 
\Lfig{bias-square}
\end{center}
\end{figure}
To make a fair comparison, the present teacher couplings for the RR graph case are all positive and constant as the square lattice case. These two distributions are very similar, implying that the bias in coupling estimates for remote spins is, even if it exists in loopy graphs, very weak for the present situation. For further quantitative information, the means of those distributions were plotted against the system size in the lower panels. Again, we observed no clear deviation from zero and no significant difference between the two cases of the square lattice and RR graph. These suggest the practicality of the theoretical results for wider situations than tree-like networks.  


\subsection{Comparison with interaction screening} \Lsec{Comparison with}
We here examine the IS cost function~\cite{vuffray2016interaction,lokhov2018optimal}, which is another common method for the inverse Ising problem. The IS cost function is given by
\be
\ell^{{\rm IS}}(x)=e^{-x}.
\ee
The quantitative comparison with the PL method is of our interest.

In \Rfig{comp_PL-IS}, we plot $\RSS$, $Q$, and $\Bias$ against $\alpha$ for the RR graph at $(c,\str)=(3,0.4)$, with two theoretical curves of IS (dashed line) and PL (dotted line). Numerical results (asterisks) are also shown to validate the theoretical result of the IS case. 
\begin{figure}[htbp]
\begin{center}
\includegraphics[width=0.32\columnwidth]{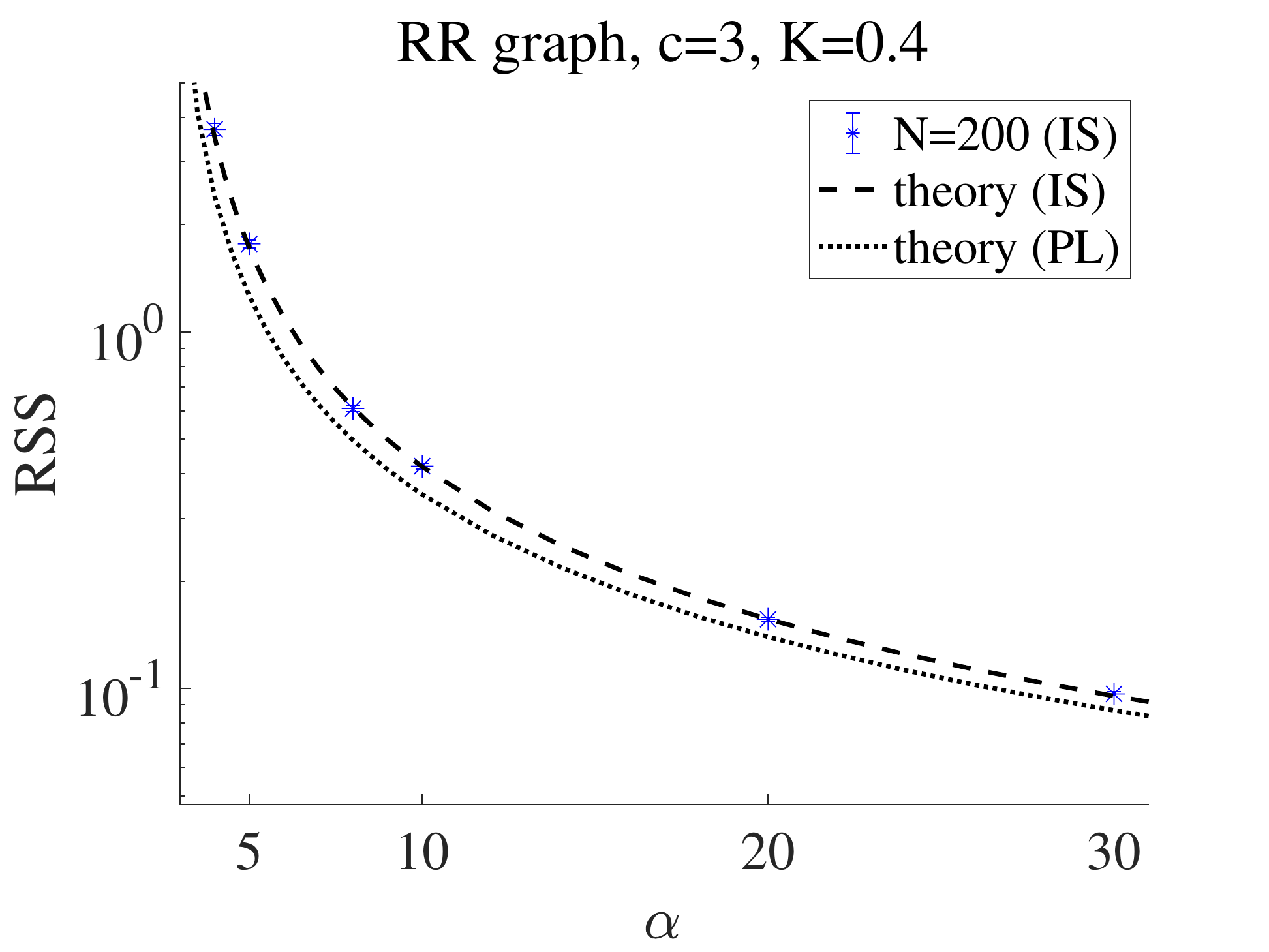}
\includegraphics[width=0.32\columnwidth]{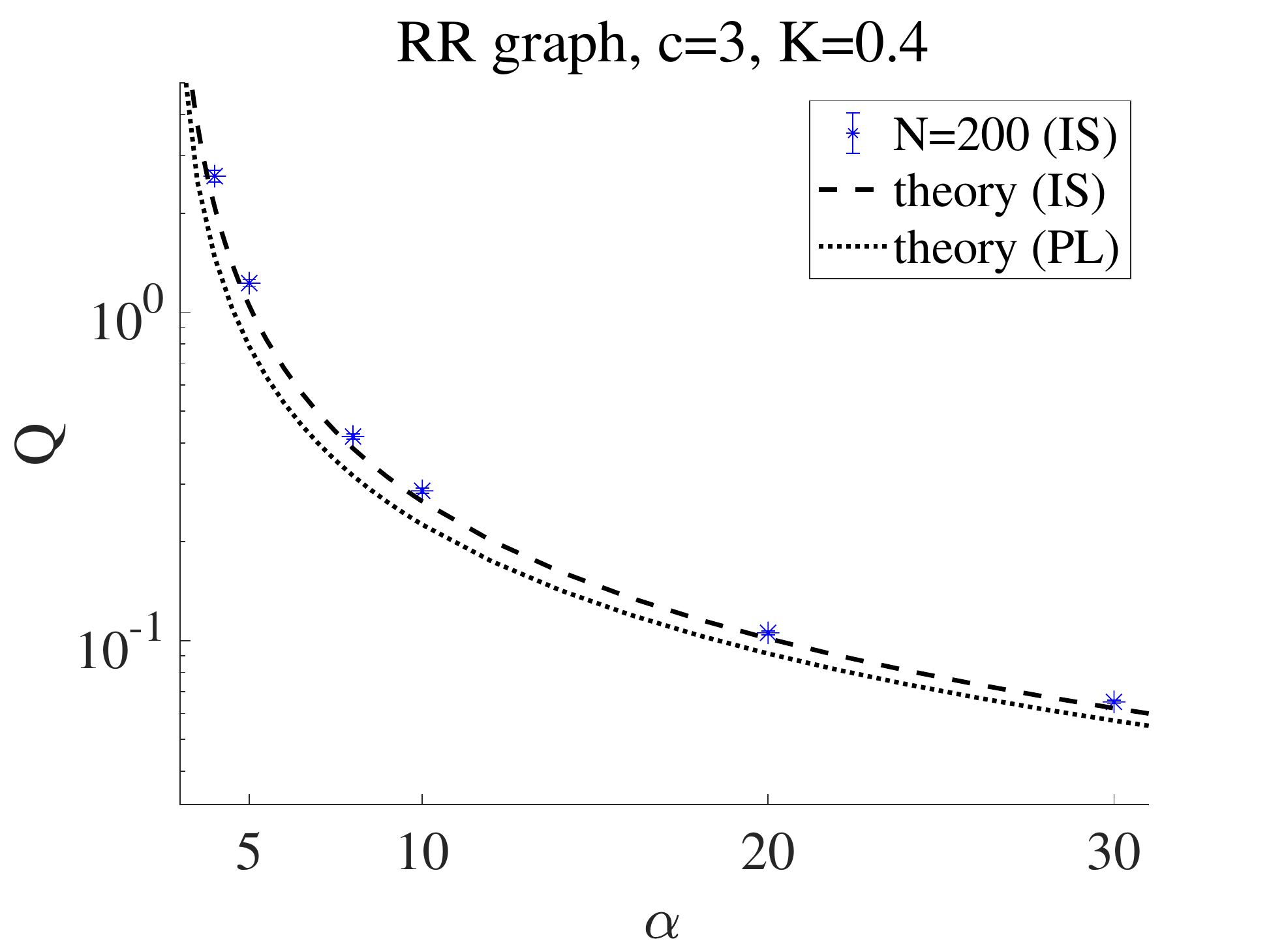}
\includegraphics[width=0.32\columnwidth]{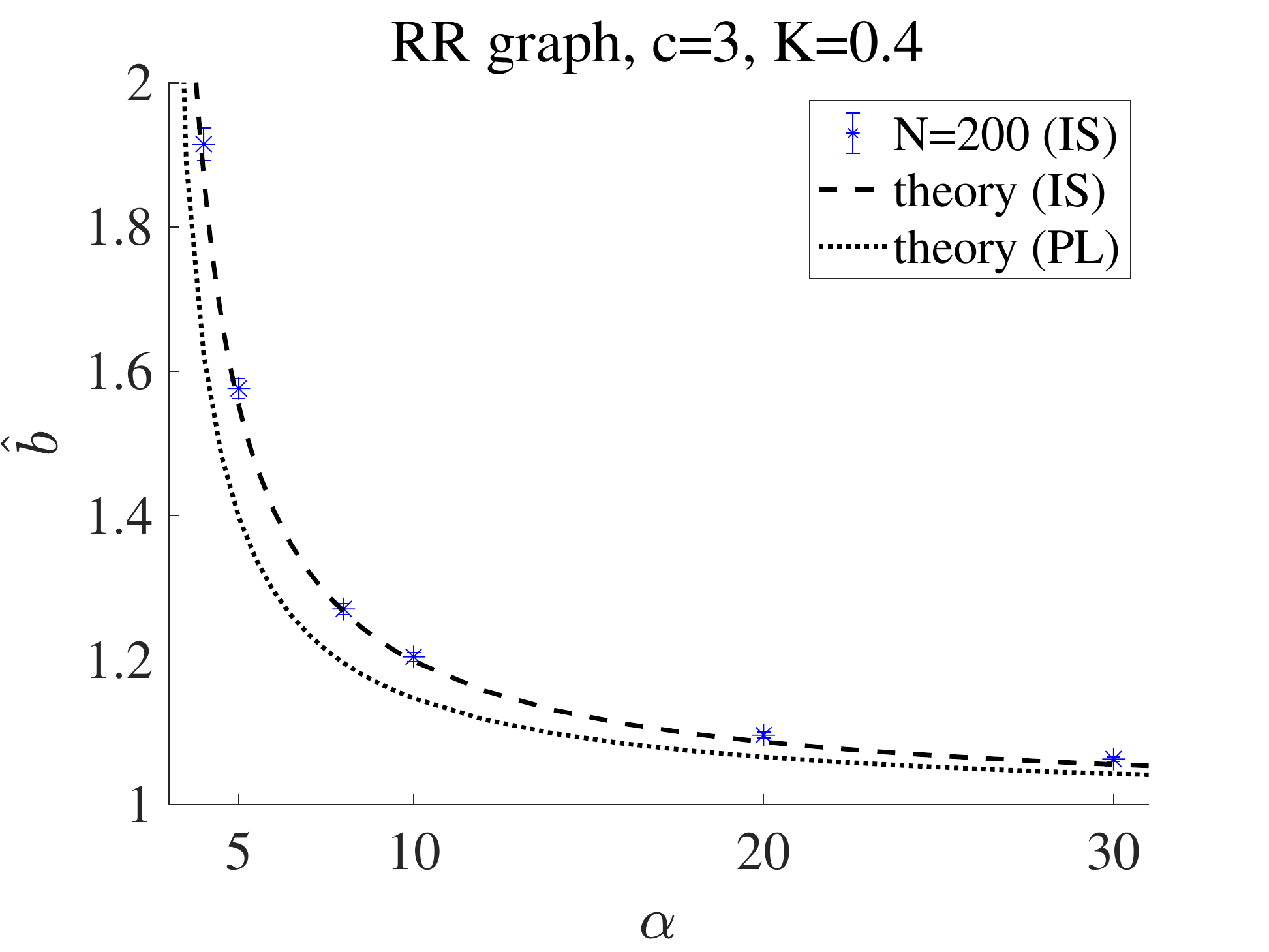}
\caption{Comparison of the PL and IS results. Plots of $\RSS$ (left), $Q$ (middle), and $\Bias$ (right) against $\alpha$ at $(c,\str)=(3,0.4)$ for the RR graph are given. The theoretical curves for PL and IS are described by dotted and dashed lines, respectively. The color markers are the numerical results for the IS case at $N=200$, whose error bars are obtained from $N_{\rm set}=100$ datasets. The consistency of the numerical and theoretical results is fairly good. The IS result provides larger values in all these three quantities than those of PL, implying that IS gives larger error, variance, and bias. }
\Lfig{comp_PL-IS}
\end{center}
\end{figure}
The important observation is that the IS result consistently gives larger values of $\RSS$, $Q$, $\Bias$ than those of PL. This implies that the IS method gives larger error, variance, and bias than PL. Although it is known that IS achieves the optimal scaling of the dataset size $M$~\cite{vuffray2016interaction,lokhov2018optimal,santhanam2012information} necessary for perfect reconstruction of the coupling network structure, it does not necessarily mean the better accuracy in terms of variance/bias. The present analysis demonstrates this in a concise way.

\section{Summary and discussion} \Lsec{Summary}
We proposed a theory to evaluate the reconstruction performance in inverse Ising problems with sparse couplings by maximizing of the pseudolikelihood in the thermodynamic limit. A large part of the theory relies on the statistical mechanical formulation in~\cite{bachschmid2017statistical}, but we refined the theoretical treatment in the cavity method to handle the teacher model with sparse couplings. The resultant expression requires a full functional form of the cavity field distribution, which is far from Gaussian but was obtained by appropriate consideration of the direct problem counterpart. The theoretical result shows  fairly good agreement with numerical experiments conducted on the RR and ER graphs, justifying our theoretical treatment. This agreement holds even for the case of the square lattice, suggesting the practicality of the present result as an approximation for loopy graphs.

The crucial assumptions of our treatment are the asymptotic behavior of the estimator \NReq{key ansatz1} and the paramagnet assumption of the teacher model, leading to the decoupled distributions of the cavity fields. The former assumption implies that the teacher's couplings can be reconstructed by the student almost perfectly, as discussed at the end of \Rsec{RR graph} according to the hypothesis testing framework, providing a theoretical reasoning to use the inverse Ising framework. The latter assumption requires the smallness of the coupling strength, implying that strongly-correlated datasets cannot be treated by the proposed theory. It will be a challenge to overcome this applicability limitation.

Concerning to the perfect reconstruction of the sparse network in the inverse Ising framework, earlier studies reported similar results in empirical and theoretical ways~\cite{decelle2014pseudolikelihood,santhanam2012information,wainwright2008graphical,ravikumar2010high}. Especially, a series of analyses by Wainwright and the collaborators~\cite{santhanam2012information,wainwright2008graphical,ravikumar2010high} derived the necessary and sufficient conditions for the perfect reconstruction in the high-dimensional limit, clarifying that the necessary size of the dataset scales as $M=O(\log N)$ when the maximum degree of the network is bounded from above. Compared to this scaling, our result on the scaling $M=O(N)$ is rather conservative. Our formulation, however, has some nontrivial advantages by deriving more detailed information about the system. For example, our formulation can treat the ER graph, the maximum degree of which is not bounded and the proofs established in~\cite{santhanam2012information,wainwright2008graphical,ravikumar2010high} are not applicable. Our result also clarifies by directly assessing the estimator's fluctuation that the hypothesis testing can actually achieve the perfect reconstruction, which provides another efficient way of reconstruction than the $\ell_1$ regularization used in the above earlier studies. The explicit computation of the bias on the estimator is also another profit. In this way, the present formulation can provide finer information in concrete cases and will be beneficial in wider situations.

For handling real-world datasets, finite magnetizations as well as possible loop structures in the network should be taken into account. For such realistic situations, the computation of $(1/N)\Tr{}C^{-1}$ and $P_{\rm cav}(h^*|\V{J}^*)$ will be more complicated. For evaluating those quantities, advanced techniques such as Bethe approximation~\cite{ricci2012bethe,nguyen2012bethe}, high-temperature expansion, and MC samplings will be useful. The ansatz \NReq{key ansatz1} should be also modified for the case of loopy graphs, as discussed in \Rsec{Applicable range}. The presented result can be still practical as an approximation for treating such situations, as demonstrated in \Rsec{Square lattice}. Certain data analysis utilizing these theoretical results will be interesting and useful.

A clear drawback of the estimator treated in this paper is that it is not informative in the region $\alpha \leq 2$, as indicated by the divergent RSS in the limit $\alpha \to 2+0$ shown in \Rsec{Numerical} and~\cite{bachschmid2017statistical}. To overcome this, the use of regularizations will be promising. The $\ell_1$ regularization will be particularly useful to control false positives in the estimated couplings. It is also possible to employ hypothesis testing in conjunction with $\ell_2$ regularization~\cite{xu2018inverse}. An extension of the present analysis to these cases is interesting and will be our focus in future work.

Another interesting extension of the present analysis might be the analysis of the model-mismatched cases where the student model cannot be equal to the teacher one. Even in such cases, some limited information in the teacher, such as the coupling network structure, might be recovered in some conditions~\cite{terada2018objective2}. Pursuing this possibility will be interesting because it can provide a better justification for applications of the inverse Ising framework to the analysis of real-world datasets. The analysis of model-mismatched cases may also have a connection to ``criticality'' observed in earlier studies~\cite{mastromatteo2011criticality}. 

The inverse Ising problem or Boltzmann machine has been treated in this paper. Although this model is much simpler than the current models of machine learning communities, we believe that it is important to enhance theoretical knowledge on such simple models to maintain the reliability and  interpretability of the results given by machine learning technologies. We hope that our study contributes to this direction, which will lead to a better understanding of and relationship with more complex models.

\section*{Acknowledgement}
This work was supported by JSPS KAKENHI Nos. 25120013 and 17H00764, and JST CREST Grant Number JPMJCR1912, Japan. 
\appendix
\section{Computations for $L$ and $S$}\Lsec{Computations}
For computing $L$, the following decomposition of the cavity fields becomes useful:
\be 
&&
h^*=\sqrt{Q^*-\frac{m^2}{q}}v_*+\sqrt{\frac{m^2}{q}}z,
\\ &&
h^a=\sqrt{Q-q}v_a+\sqrt{q}z,
\ee
where $v_a,v_*,z$ are i.i.d Gaussian variables with zero mean and unit variance. It is easy to confirm that this decomposition reproduces the covariances among $\{ h^*,h^1,\cdots,h^n \}$. Using this and performing the integration with respect to $v_*$, we get 
\be
L(Q^*,Q,q,m)=\int Dz~e^{\sqrt{\frac{m^2}{q}}z-\frac{1}{2}\frac{m^2}{q}}
\lb
\int Dv~e^{-\IT \ell \lb \sqrt{Q-q}v+\sqrt{q}z \rb}
\rb^n,
\Leq{L_fc}
\ee
where we use the relation $Z_0=2e^{\frac{1}{2}Q^*}$, which was canceled with a factor appearing by the integration of $v_*$. \BReq{L_fc_nzero} is easily derived from this.

For computing the entropic term $S(C^{\bs 0},\V{J}^*,Q,q,m)$, we use the rescaled variable $\V{W}=\sqrt{N}\V{J}$ and set the integration measure as $\Tr{\V{J}}=\int d\V{W}$. Here, we use the uniform measure because in the present setting the student has no prior information about the teacher couplings. If certain prior knowledge is available such as the teacher coupling sparseness, it can be suitable to introduce another measure such as Laplace prior.  Further, we represent the delta functions by the Fourier expressions as follows:
\subbe
\Leq{subshell}
\be
&&
\delta \lb Q-\frac{1}{N}\sum_{i,j}C^{\bs 0}_{ij}W^a_{i}W_{j}^a \rb
=C_1\int d\tilde{Q}~e^{
\frac{1}{2}N\tilde{Q}Q-\frac{1}{2}\tilde{Q}\sum_{i,i}C^{\bs 0}_{ij}W^a_{i}W_{j}^a
},
\Leq{Q-subshell}
\\ &&
\delta \lb q-\frac{1}{N}\sum_{i,j}C^{\bs 0}_{ij}W^a_{i}W_{j}^b \rb
=C_2\int d\tilde{q}~e^{
-N\tilde{q}q+\tilde{q}\sum_{i,j}C^{\bs 0}_{ij}W^a_{i}W_{j}^b
},
\Leq{q-subshell}
\\ &&
\delta \lb m-\frac{1}{N}\sum_{i,j} C^{\bs 0}_{ij}W^*_{i}W_{j}^a\rb
=C_2\int d\tilde{m}~e^{
-N\tilde{m}m+\tilde{m}\sum_{i,j} C^{\bs 0}_{ij}W^*_{i}W_{j}^a
},
\Leq{m-subshell}
\ee
\subee
where the integration contour is the imaginary axis and $C_1,C_2$ are appropriate normalization constants; however, these points are irrelevant and ignored hereafter. Inserting \Leq{subshell} this into \Req{S_def}, we get
\be
e^{NS}=\int d\T{Q}d\T{q}d\T{m}~e^{S_{X}}
\int \prod_a d\V{W}^a~e^{U},
\ee
where
\be
&&
\mc{S}_{X}=N\lb \frac{1}{2}n\hat{Q}Q-\frac{1}{2}n(n-1)\T{q}q-n\T{m}m \rb,
\\ &&
U
=
-\frac{1}{2}\T{Q}\sum_{a}\lb \V{W}^a  \rb^{\top} C^{\bs 0}\V{W}^a 
+\T{q}\sum_{a<b}\lb \V{W}^a  \rb^{\top} C^{\bs 0}\V{W}^b
+\T{m}\sum_{a}\lb \V{J}^* \rb^{\top} C^{\bs 0}\V{W}^a 
\no \\ &&
=
-\frac{1}{2}(\T{Q}+\T{q})\sum_{a}\lb \V{W}^a  \rb^{\top} C^{\bs 0}\V{W}^a 
+\frac{1}{2}\T{q}\sum_{a,b}\lb \V{W}^a  \rb^{\top} C^{\bs 0}\V{W}^b
+\T{m}\sum_{a}\lb \V{J}^* \rb^{\top} C^{\bs 0}\V{W}^a .
\Leq{U}
\ee
To decouple different replicas and components of $\{\V{W}^a \}_a$, we use the expression $C^{\bs 0}=O^{\top}\Lambda O$, where $\Lambda$ is the diagonal matrix consisting of the eigenvalues $\{\lambda_i\}_i$ and $O$ is the appropriate orthogonal matrix. Performing the variable transformation $\T{\V{W}}=O\V{W}$ and applying the Hubbard--Stratonovich transformation, we get 
\be
&&
\int \prod_a d\V{W}^a~e^{U}
=
\int \prod_{i}Dz_i
\int \prod_a d\T{\V{W}}^a~
e^{
-\frac{1}{2}(\T{Q}+\T{q})\sum_{a}\sum_{i}\lambda_{i}\lb \T{W}^a_{i}\rb^2
+\sum_{a}\sum_{i} \lb \lambda_i\T{W}^{*}_i\T{m}+\sqrt{\lambda_i \T{q} }z_i \rb \T{W}^a_{i} 
}
\no \\ &&
=
\int \prod_{i}Dz_i~
e^{n\sum_{i}\lbb \frac{1}{2}\frac{\lb \sqrt{\lambda_i}\T{W}^{*}_i\T{m}+\sqrt{\T{q} }z_i \rb^2 }{\lb \T{Q}+\T{q} \rb} +\frac{1}{2}\lb \log 2\pi - \log \lambda_i -\log(\T{Q}+\T{q}) \rb\rbb}
\no \\ &&
=
e^{
-\frac{N}{2} \log \lb 1-\frac{n\T{q}}{\T{Q}+\T{q} }  \rb
+
\frac{1}{2} n\sum_{i}\frac{\lambda_i\lb \T{W}^*_{i}\rb^2\T{m}^2}{\T{Q}+\T{q}(1-n)} 
+
\frac{n}{2}\sum_i\lb \log 2\pi - \log \lambda_i -\log(\T{Q}+\T{q}) \rb
}
\equiv e^{\mc{S}_J}.
\ee
Note that this quadratic form with respect to $\T{\V{W}}$ implies that $\T{\V{W}}$ essentially obeys Gaussian, and thus the estimator $\hat{\V{J}}$ also does. This knowledge of the distribution can be used for  hypothesis testing as addressed in the main text.

In the thermodynamic limit $N\to \infty$, we can use the saddle-point (or Laplace) method to avoid the explicit integrations with respect to $\T{Q},\T{q},\T{m}$. This yields
\be
&&
S=\Extr{\T{Q},\T{q},\T{m}}\lbb \frac{\mc{S}_{X}+\mc{S}_{J}}{N}\rbb=\Extr{\T{Q},\T{q},\T{m}}
\Bigg\{
\frac{1}{2}n\T{Q}Q-\frac{1}{2}n(n-1)\T{q}q-n\T{m}m
-\frac{1}{2} \log \lb 1-\frac{n\T{q}}{\T{Q}+\T{q} } \rb 
\no \\ &&
+\frac{1}{2} n\frac{Q^*\T{m}^2}{\T{Q}+\T{q}(1-n)} 
+
\frac{n}{2}\lb \log 2\pi-\log(\T{Q}+\T{q}) \rb
 -\frac{n}{2N}\Tr{} \log C^{\bs 0} 
\Bigg\}.
\ee
where we used the relations $\sum_{i}\lambda_i\lb \T{W}^*_{i}\rb^2=NQ^*,~\sum_{i}\log\lambda_i=\Tr{}\log C^{\bs 0}$. The limit $n\to 0$ leads to
\be
&&
\lim_{n\to 0}\frac{S}{n}=\Extr{\T{Q},\T{q},\T{m}}
\Bigg\{
\frac{1}{2}\T{Q}Q+\frac{1}{2}\T{q}q-\T{m}m
+\frac{1}{2}\frac{\T{q}+Q^*\T{m}^2}{\T{Q}+\T{q}} 
\no \\ &&
+
\frac{1}{2}\lb \log 2\pi-\log(\T{Q}+\T{q}) \rb
 -\frac{1}{2N}\Tr{} \log C^{\bs 0} 
\Bigg\},
\Leq{S_fc_nzero_pre}
\ee
and the extremization condition gives
\be
\T{Q}=\frac{Q-2q+m^2/Q^*}{(Q-q)^2},~
\T{q}=\frac{q-m^2/Q^*}{(Q-q)^2},~
\T{m}=\frac{m/Q^*}{Q-q}.
\Leq{tilde parameters}
\ee
Substituting these relations into \Req{S_fc_nzero_pre}, we obtain \Req{S_fc_nzero}. If we ignore the terms related to $m$ and $\T{m}$, we have \Req{S_sparse_nzero}.

\section{Derivation of macroscopic parameters $R$ and $\rho$}\Lsec{Derivation}
To derive the expressions of $R$ and $\rho$, we can employ the technique of auxiliary variables. We introduce two terms $h_{R}\sum_a \lb \V{W}^{a}\rb^{\top}\V{W}^{a}$ and $h_{\rho}\sum_{a}\lb \V{W}^{*}\rb^{\top}\V{W}^{a}$ in \Req{U}, and perform the same line of computations as in \Rsec{Computations}. As a result, the entropic term is modified to the following expression:
\be
&&
\lim_{n\to 0}\frac{S}{n}=\Extr{\T{Q},\T{q},\T{m}}
\Bigg\{
\frac{1}{2}\T{Q}Q+\frac{1}{2}\T{q}q-\T{m}m
\no \\ &&
+\frac{1}{2N}\sum_{i}
\lb
\frac{
(\T{m}\lambda_i+h_{\rho})^2\lb \T{W}^*_{i}\rb^2+\lambda_i\T{q}   
}{
(\T{Q}+\T{q})\lambda_i-2h_R
} 
+
\log 2\pi 
-\log \lb (\T{Q}+\T{q})(\lambda_i-2h_{R}) \rb 
\rb
\Bigg\}.
\Leq{S_fc_nzero_aux}
\ee
Taking the differentiation with respect to $h_{\rho}$ and taking the limit $h_{\rho},h_{R}\to 0$, we get
\be
\rho=
\lim_{h_{\rho},h_{R}\to 0}\Part{}{h_{\rho}}{}\lim_{n\to 0}\frac{S}{n}
=
\frac{1}{N}\sum_{i}\frac{\T{m}\lb \T{W}^*_{i} \rb^2}{\T{Q}+\T{q}}
=\frac{m}{Q^*}R^*.
\ee
The last expression is obtained by using \Req{tilde parameters}. Similarly, 
\be
&&
R
=
\lim_{h_{\rho},h_{R}\to 0}\Part{}{h_{R}}{}\lim_{n\to 0}\frac{S}{n}
=
\frac{1}{N}\sum_{i}
\lb
\frac{\T{m}^2 \lb \T{W}^*_{i} \rb^2}{\lb \T{Q}+\T{q}\rb^2}
+
\frac{ \T{Q}+2\T{q}}{\lb \T{Q}+\T{q}\rb^2}\frac{1}{\lambda_i}
\rb
\no \\ &&
=
\lb \frac{m}{Q^*}\rb^2R^*
+
\lb Q-\frac{ m^2}{Q^*}\rb 
\frac{1}{N}\Tr{}\lb C^{\bs 0}\rb^{-1}.
\Leq{R}
\ee
These give \Req{R and rho}.

In the sparse case, we need to compute $\sum_{i\in \bar{\Omega}}\Delta_i^2$ for computing the RSS. By construction, this is equivalent with $R$ when $m$ is absent. Hence $\sum_{i\in \bar{\Omega}}\Delta_i^2$ is given by putting $m=0$ in \Req{R}, leading to \Req{RSS_sparse2}.

\section{Details of numerical experiments}\Lsec{Details of numerical}
The actual experimental procedures are summarized as follows. We first generated a random graph and the teacher couplings on it, and obtained spin snapshots using MC sampling. Then, we randomly chose a center spin $s_0$ from the whole spins and learned the couplings connected to $s_0$ by minimizing the PL cost function defined with a dataset obtained from the sampled spin configurations. This single sequence of operations provided single values of the quantities of interest, such as $\RSS$ and $Q$. To obtain the error bars of those quantities, we repeated this sequence many times. Here, the experiment had three different sources of fluctuations: the generated teacher model (graph shape and couplings), the choice of the center spin, and the MC sampling. We did not discriminate between these three fluctuations unless explicitly mentioned, and we defined the error bar as the standard error among the obtained values according to their recurrence; the number of datasets obtained this way is hereafter denoted as $N_{\rm set}$. In the MC sampling, we started from a random initial configuration and updated the state by the standard Metropolis method; one MC step (MCS) is defined by $N$ trial flips of spins, where $N$ is the total number of spins. We discarded the first $10^{5}$ MCSs as burn-in to avoid systematic errors from the initialization. Furthermore, to avoid possible correlations in samples, each dataset for learning was generated by subsampling from a much larger dataset, which consists of all the configurations recorded after every few numbers of MCS. The size of the subsampled dataset was chosen to be at least five times smaller than that of the larger dataset. The optimization algorithm is a standard trust-region method using the second-order expansion of the objective function. 

\bibliographystyle{unsrt}
\bibliography{testmaxent,obuchi}

\end{document}